\documentclass[10pt,journal,compsoc]{IEEEtran}
\ifCLASSOPTIONcompsoc
  \usepackage{cite}
  
\else
  \usepackage{cite}
\fi
\usepackage{amsmath,amsfonts,amssymb}
\usepackage{algorithmic}
\usepackage{algorithm}
\usepackage{array}
\usepackage[caption=false,font=normalsize,labelfont=sf,textfont=sf]{subfig}
\usepackage{textcomp}
\usepackage{stfloats}
\usepackage{url}
\usepackage{verbatim}
\usepackage{graphicx}
\usepackage{cite}
\usepackage{mathrsfs}
\newtheorem{definition}{Definition}

\usepackage{multicol}
\usepackage{multirow}
\usepackage{booktabs}
\usepackage{makecell}
\usepackage{MnSymbol}
\usepackage{xcolor}
\usepackage{bbm}
\usepackage{float}
\usepackage{footmisc}

\begin{document}

\title{Causal Longitudinal Image Synthesis}

\author{Yujia Li, Han Li, ans S. Kevin Zhou,~\IEEEmembership{Fellow,~IEEE,}
}



\IEEEtitleabstractindextext{%
\begin{abstract}
Clinical decision-making relies heavily on causal reasoning and longitudinal analysis.
For example, for a patient with Alzheimer's disease (AD), how will the brain grey matter atrophy in a year if intervened on the A-beta level in cerebrospinal fluid? 
The answer is fundamental to diagnosis and follow-up treatment. However, this kind of inquiry involves counterfactual medical images which can not be acquired by instrumental or correlation-based image synthesis models. Yet, such queries require counterfactual medical images, not obtainable through standard image synthesis models. Hence, a causal longitudinal image synthesis (CLIS) method, enabling the synthesis of such images, is highly valuable.
However, building a CLIS model confronts three primary yet unmet challenges: mismatched dimensionality between high-dimensional images and low-dimensional tabular variables, inconsistent collection intervals of follow-up data, and inadequate causal modeling capability of existing causal graph methods for image data. In this paper, we established a tabular-visual causal graph (TVCG) for CLIS overcoming these challenges through a novel integration of generative imaging, continuous-time modeling, and structural causal models combined with a neural network. Specifically, we first depict the causality between tabular variables including demographic variables, clinical biomarkers, and brain volume size via a tabular-only causal graph (TOCG), and then further establish a tabular-visual causal graph (TVCG) to causally synthesize the brain MRI by developing an intervened MRI synthesis module (ISM) as an edge between TOCG and MRI.
We train our CLIS based on the ADNI dataset and evaluate it on two other AD datasets, which illustrate the outstanding yet controllable quality of the synthesized images and the contributions of synthesized MRI to the characterization of AD progression, substantiating the reliability and utility in clinics.
\end{abstract}}


\maketitle
\IEEEpeerreviewmaketitle

\section{Introduction}
\label{sec:introduction}

\IEEEPARstart{C}{linical} decision-making heavily depends on both \textbf{longitudinal} pattern comparison and \textbf{causal reasoning}~\cite{causal_medical_diagnosis, causal_diagnose, medical_consultation, Interactive_Diagnosis}.  For instance, in clinical decision-making related to Alzheimer's disease (AD), the most common type of dementia and a neurodegenerative disorder with a progression that can extend over decades~\cite{longitudinal_AD1, longitudinal_AD2, longitudinal_AD3}, physicians initially assess changes in brain structure over time through longitudinal magnetic resonance image (MRI) scans~\cite{causal_ad_mri_gen}. They then integrate these changes with other tabular-variable factors, such as age and cognitive ability, drawing on their medical experience to understand the progression of AD.
In this context, conducting longitudinal pattern comparison is relatively straightforward if longitudinal data are available, but integrating brain structure changes with other tabular-variable factors to understand the progression of AD poses a significant challenge. This difficulty arises because these factors are not impacted independently but are causally interrelated. Years of clinical experience in AD are essential for physicians to develop their own causal reasoning mechanism to understand AD progression.

\begin{figure}[t]
  \centering
  \includegraphics[width=0.48\textwidth]{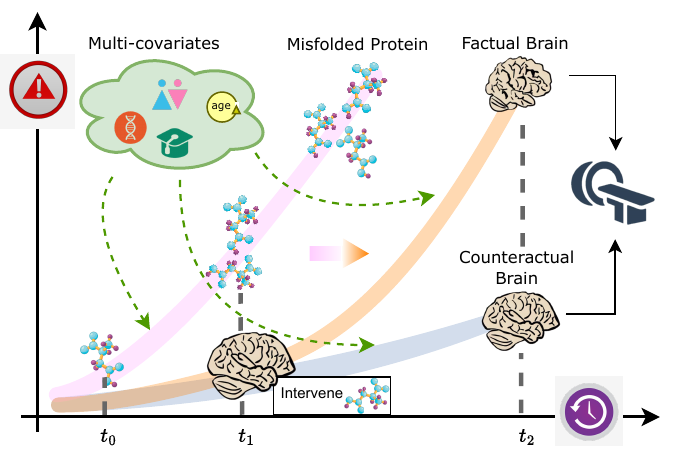}
  \caption{The illustration of causal and longitudinal disease progression. Along the time axis, the accumulation of misfolded proteins has a causal effect on brain atrophy. Multiple covariates, including gene type, sex, education level, and baseline age, also have causal effects on the disease trajectories (shown in yellow). The blue trajectory represents the \textbf{counterfactual} brain atrophy when successfully intervened on the misfolded protein level at $t_1$, which deviates from the fact and results in a different MRI at $t_2$. We attempt to develop a computational model that leads to such a success.}
  \label{fig:intro}
\end{figure}

However, even with high-quality longitudinal MRI and a robust causal reasoning mechanism, comprehending AD's progression remains complex. This complexity largely stems from the need for physicians to consider specific `If' questions to understand AD progression in fine granularity, such as: \textit{`If the patient were 10 years younger, how would their brain image appear?'} or \textit{'If there is an increase in the  A-beta level in cerebrospinal fluid (CSF), how would it impact the grey matter volume (GMV)?'} These inquiries, known as \textit{counterfactual} questions~\cite{pearl_2009, causal_ad_mri_gen}, involve hypothesizing specific interventions that diverge from actual scenarios. As in Fig.~\ref{fig:intro}, brain atrophy, a key clinical indicator in AD diagnosis, is influenced by numerous factors, including multiple covariates and interventions. Typically, physicians compare brain size changes between MRI scans acquired say at $t_1$ and $t_2$. They then attempt to answer counterfactual questions at different time points, like $t_0$ or beyond $t_2$, based on the extrapolation of these changes and other factors. Subsequently, they determine the timing and nature of interventions needed to potentially slow down or reverse brain atrophy, which is indicated by the blue line in Fig.~\ref{fig:intro}. 

Addressing this issue ideally requires an imaging modality that can directly answer these counterfactual questions or assist physicians in understanding the causal mechanisms of AD. Unfortunately, no existing imaging modalities can meet this need. Consequently, a computational method, that generates novel longitudinal synthetic images while incorporating AD's causal mechanisms and providing insights into counterfactual questions, emerges as a preferable solution. To this end, we hereby attempt to establish a tabular-visual causal graph (TVCG) for \textbf{causal longitudinal image synthesis} (CLIS) task that incorporates both causality and longitudinality, with a novel integration of causal graph ~\cite{pearl_2009} and generative imaging.

Specifically, we first depict the causality between longitudinal tabular variables including longitudinal demographic variables, clinical biomarkers, and brain volume size via a tabular-only causal graph (TOCG), and then further establish a tabular-visual causal graph (TVCG) by using an intervened MRI synthesis module (ISM) as an edge between longitudinal image and TOCG. This framework allows for the easy synthesis of counterfactual MRIs, addressing the aforementioned counterfactual questions through simple modifications of the variables (interventions) in TVCG.

When developing our TVCG, we encounter three primary yet unmet challenges: (1) {\it Mismatched dimensionality}: While the causal graph has demonstrated success in fields such as epidemiology, econometrics, and medicine ~\cite{epidemiology,econometrics,medicine}, it primarily deals with low-dimensional tabular variables. In contrast, causal image synthesis involves both high-dimensional images and low-dimensional tabular variables. 
(2) {\it Inconsistent collection intervals} of training data: The most direct way to model AD progression is through training models with longitudinal data. Yet, the acquisition time of existing longitudinal MRI data varies from patient to patient, making it difficult to develop a model that can accommodate this variability.
(3) {\it Inadequate causal modeling capability}: The causal mechanism of AD involves intricate interactions among various factors. Traditional causal graph methods, like Structural Causal Models (SCM), require predefined graph structures and often rely on linear causal relationship modeling (i.e., edge)~\cite{Linear_causal_modeling}, which can struggle to accurately represent these intricate relationships. 


To address the challenge of {\it Mismatched dimensionality}, we propose the generative imaging method, ISM, as a feasible solution. The key idea of IMS is to represent a high-dimensional 3D medical volume with a medium-dimensional latent vector, which serves as a bridge between low-dimensional tabular variables and high-dimensional medical images. In ISM, we train a 3D StyleGAN~\cite{karras2019style} as a latent-to-image generator that defines a latent-to-image mapping function, and an accompanying neural network (NN) as an image-to-latent decoder that defines an image-to-latent mapping function. 
To address the challenge of {\it Inconsistent longitudinal collection intervals} of training data, we propose to incorporate the time interval as an independent variable in our model for the capability of continuous time prediction. Existing approaches tend to construct multiple models for a subset of discrete time points~\cite{LDGAN, multi-information-GAN, TR-GAN}, which is not scalable and suffers from a scarcity of training data. 
To address the challenge of {\it Inadequate causal modeling capability} of traditional causal graph methods, our TVCG uses a Hybrid Causal Graph design, which merges Temporal Causal Graph (TCG) with Structural Causal Model (SCM) techniques. This hybrid model incorporates TCG's causal discovery mechanism and enhances it with SCM’s soft (deterministic or probabilistic) edge-fitting methods, thus leveraging the strengths of both methods. Moreover, during SCM edge-fitting, we use a neural network (NN) based approach. Compared with the commonly used linear edge representation in SCM, the NN edge representation demonstrates enhanced capability in modeling causal relationships compared to linear methods.


In summary, our key contributions are as follows. 
\begin{itemize}
\item We propose a tabular-visual causal graph (TVCG)  designed for the {\bf causal longitudinal image synthesis (CLIS)} task. TVCG effectively tackles the three major challenges inherent in CLIS: the mismatched dimensionality of a high-dimensional image and low-dimensional tabular variables; inconsistent collection intervals of training data; and inadequate causal modeling capability of traditional causal graph methods.


\item We evaluate our TVCG in both causality and longitudinal brain MRI synthesis tasks by several metrics. Our TVCG achieves the {\bf best performances} compared with previous methods. Additionally, the synthesized images have also proven beneficial in a preliminary manner for predictive tasks such as the clinical characterization of AD. 

\item Beyond the image synthesis, TVCG shed light on the {\bf causal relationships} among demographic, bio-markers, brain volume variables, and MR images. The causality is verified on external validation cohorts and may help to alleviate the spurious correlations for machine learning models.





\end{itemize}


\section{Related work} \label{sec:related}
\subsubsection{Medical Image Synthesis}
Medical image synthesis has a potential for mitigating challenges such as limited or absent data, privacy concerns, and dataset biases. 
Image synthesis across MRI, computed tomography (CT), and positron emission tomography (PET)~\cite{MRI_to_CT_review, MR-CT-miccai20223, Multi-Modality_to_PET, MR_to_PET1, MR_to_PET2, MR_to_PET3}and among different MRI sequences~\cite{Multi-Contrast_MRI_TMI_2021, zhang2023unified, Neuroimage_Synthesis}, different resolutions~\cite{SR-review}, or the MRI's of different categories (health or disease)~\cite{cGAN_AD1, cGAN_AD2} have been widely explored. 

However, limited research focuses on the prediction of medical images at future time-points because of the challenges of processing various session intervals and capturing the fine structure alternations among the scans for a single individual. LDGAN~\cite{LDGAN} and MI-GAN~\cite{multi-information-GAN} predict multiple future data by training multiple models and are incapable of handling varying session intervals. Fan et al.~\cite{TR-GAN} propose TR-GAN to deal with input sequences of varying lengths and generate future variant sessions, but it does not consider other modalities, such as bio-markers and demographic information.

There have been some attempts to introduce causality into medical image synthesis for multiple sclerosis~\cite{MS1, MS2}, brain tumor~\cite{counterfactual_diffusion} and AD~\cite{DSCM_for_AD}. However, they concentrate on 2D images and lack effective metrics to evaluate the quality of synthesized images via causal intervention. They are also sectional and thus might be improper for downstream tasks. 
\subsubsection{Causal Model for Alzheimer's disease}
Causal relationships in AD have been widely investigated in the fields of pathophysiology. The related factors include age~\cite{age1, age2, age3}, gender~\cite{sex1, sex2, sex3}, education level~\cite{education1, education2}, risk genes\cite{AD_gene1, AD_gene2, VGF_nature}, accumulated misfolded amyloid beta ($A\beta$) and hyperphosphorylated tau protein ($p\tau$)~\cite{Abeta_and_tau1, Abeta_and_tau2, Abeta_and_tau3, Abeta_and_tau4}, grey matter loss~\cite{GM1, GM2, GM3}, and enlarged ventricles~\cite{ventricles1, ventricles3}. The current research exhibits a dual emphasis. The first group of methods concentrates on the interactions between two or three specific factors~\cite{education_and_gender, age_and_sex_and_education, age_and_APOE_and_sex, Abeta_and_age, age_and_GM, biomarkers_and_ventricles, GM_and_WM}, while the others~\cite{causal_discovery_algorithms_for_AD, causal_model_from_report} underscore the establishment of causal relationship networks among multiple factors, which is more related to our work. 

Iturria-Medina et. al.~\cite{Multifactorial_causal_model_AD} propose a multifactorial causal model to study the disease progression~\cite{vascular_causal_model_AD} and furthermore potential intervention. Shen et. al.~\cite{causal_discovery_algorithms_for_AD} apply several causal discovery algorithms to AD to examine whether they can recover the causal graph from observational clinical data. Hu et. al.~\cite{causal_model_from_report} collect prior causal knowledge and orient the arcs of a causal Bayesian network from diagnosed patient data. However, they either use extracted features of MRI or do not include MRI because of the intractability of the high-dimensional image data, which limits the modeling performance and clinical significance.

\subsubsection{Characterization of Alzheimer's disease}
A treatment given at an early stage of AD is more effective. Thus, a crucial task is to identify those at early disease stages, who are likely to progress over the short-to-medium term (1-5 years)~\cite{tadpole} and may react positive to treatment. Thus, lots of research try to forecast early AD using some key features such as clinical status, cognitive decline, and brain atrophy.

There have been multiple works using various models, such as Gaussian process~\cite{guassian_process}, recurrent neural network (RNN)~\cite{RNN}, Bayesian latent variable model~\cite{Multi-task}, and parametric model~\cite{parametric}. While all these models deal with the multi-modal data, the lack of future MRI restricts their performances; yet our method that is able to predict future MRI can alleviate this restriction.

The Alzheimer’s Disease Prediction Of Longitudinal Evolution (TADPOLE) Challenge~\cite{tadpole2} compares the prediction performance of 58 algorithms. The participants train on historical data from the Alzheimer’s Disease Neuroimaging Initiative (ADNI)~\cite{ADNI} or other accessible datasets and are required to make monthly forecasts over a period of 5 years of total volume of the ventricles. We also apply our method to this part of the challenge and show better results.

\section{Preliminaries} \label{sec:prelim}
To describe the causal relationships between time series, either Temporal Causal Graph (TCG)\cite{temporal_causal_graFph} or Structural Causal Model (SCM) \cite{pearl_2009} can be used. The edges in TCG can be dynamically modeled through training to indicate the presence or absence of causality between binary or continous variables~\cite{Causal_probabilistic_networks} in \textbf{a binary manner}. 
In contrast, in SCM, also known as Structural Equation Models (SEM), the presence or absence of each edge is predefined. These edges are then trained to represent causal relationships through either \textbf{deterministic or probabilistic} functions. In the following, we will introduce them, respectively.
\subsection{Temporal Causal Graph (TCG)}\label{sec:Preliminary Window Causal Graph}
We use a window causal graph, a special type of temporal causal graph, to introduce the mechanism of the causal relationships between the variables in TCG. A window causal graph $\mathscr{G} = (\boldsymbol{V},\boldsymbol{E})$ is a direct acyclic graph (DAG), defined as follows,
\begin{definition}[Window Causal Graph]\label{def:window causal graph}

Let $\boldsymbol{x}$ be a $d$ dimension variable which contains $d$ components $\boldsymbol{x}_1, ... ,\boldsymbol{x}_d$. Each component (e.g. $\boldsymbol{x}_1$) contains a value to represent an attribute.  In Window Causal Graph  $\mathscr{G} = (\boldsymbol{V},\boldsymbol{E})$, we got two sets: Node set $\boldsymbol{V}$ and edge set $\boldsymbol{E}$. The node set $\boldsymbol{V}$ contains various $\boldsymbol{x}$ in different timestep $t_0,t_1, ... ,t_{\tau}$, Each $\boldsymbol{x}$ still contains $d$ components:
\begin{equation}
    V=[x^1,x^2, ... ,x^t], x^t=[{x}_1^t,x_2^t ... ,{x}_d^t].
\end{equation}

As for the Edge set $E$, we employ lag-specific DIRECTED links to interconnect different components, represented as $\boldsymbol{x}_i^{t_i}\rightarrow \boldsymbol{x}_j^{t_j}$, $\boldsymbol{x}_i^{t_i} \neq \boldsymbol{x}_j^{t_j}$. These links can either be Forward-Cross-Time link ($t_i<t_j$) or within-time link ($t_i=t_j$), However, Reverse-Cross-Time link ($t_i>t_j$) are not permitted. The difference  $t_j-t_i$ is termed as a time lag $\tau$.




\end{definition}

\begin{figure}[!t]
\centering
\subfloat[\scriptsize{window causal graph}]{\includegraphics[width=0.35\linewidth]{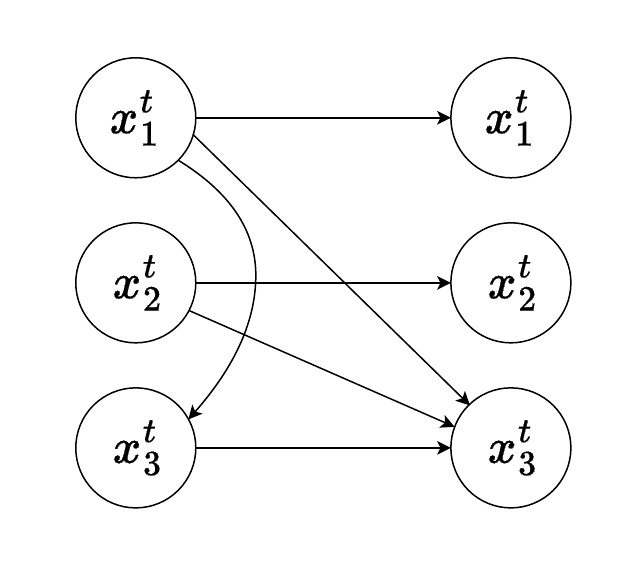}%
\label{Window Causal Graph}}
\hfil
\subfloat[\scriptsize{summary causal graph}]{\includegraphics[width=0.35\linewidth]{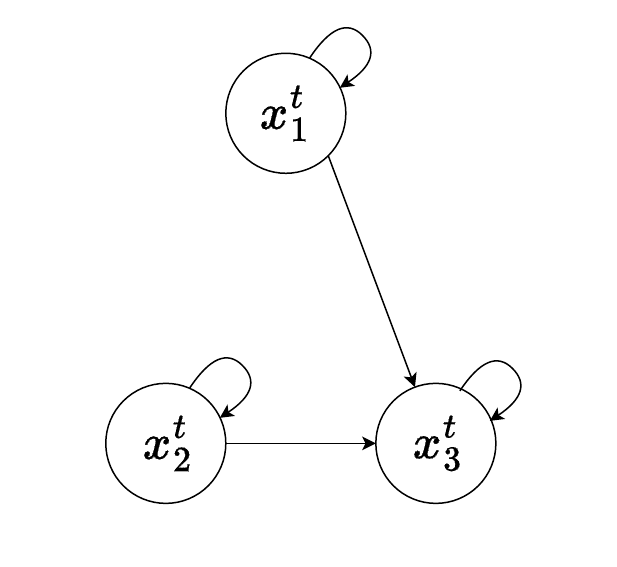}%
\label{Summary Causal Graph}}
\caption{An illustration of (a) a window causal graph of three variables in two-time steps. (b) the corresponding deduced summary causal graph.}
\label{Temporal Causal Graph}
\end{figure}

Fig.~\ref{Window Causal Graph} gives an illustration of a window causal graph. We set the window size as 1, hence the graph only focuses on two timesteps, $t$ and $t+1$. This graph represents: $\boldsymbol{x}_1$ causes $\boldsymbol{x}_2$ with a time lag of $1$, $\boldsymbol{x}_2$ causes $\boldsymbol{x}_3$ with a time lag of $1$, $\boldsymbol{x}_1$ causes $\boldsymbol{x}_3$ with a time lag of both $0$ and $1$, and $\boldsymbol{x}_1, \boldsymbol{x}_2, \boldsymbol{x}_3$ all cause itself with a lag of 1. To be more concise, a simplified version of the window causal graph, termed as summary causal graph, is proposed to simplify the representation, which is shown in in Fig.~\ref{Summary Causal Graph}. 

Subsequently, we use a summary causal graph as our representation.

\begin{definition}[Causal discovery]\label{sec:Preliminary Causal Discovery}
Causal discovery refers to recovering a causal graph (TCG in this work) from observational data~\cite{Discovering_Temporal_Sequence_Patterns}. There have been multiple causal discovery methods, mainly including constraint-based\cite{FCI}, score-based\cite{GES}, and functional-causal-models-based\cite{DirectLiNGAM}. Most of the methods are designed based on multiple assumptions, primarily including the Causal Sufficiency and Faithfulness assumptions:

\newtheorem{assumption}{Assumption}
\begin{assumption}[Causal Sufficiency]\label{Causal Sufficiency Assumption}
A set of variables is causally sufficient if all common causes of all variables are observed. 
\end{assumption}

\begin{assumption}[Faithfulness]\label{Faithfulness}
A graph $\mathscr{G}$ and a compatible probability distribution $P$ are faithful to one another if all and only the conditional independence relations true in $P$ are entailed by the Markov condition applied to $\mathscr{G}$.
\end{assumption}

After the causal discovery is finished, an effective way to test its validness is Partial correlation analysis. Specifically, under the Causal Sufficiency and Faithfulness assumptions, for the observed variable set $\mathcal{X} = \{ x_1, x_2, ..., x_N \}$
\begin{equation}
  x_i \nupmodels x_j | \hat{\mathcal{X}} \iff x_i \leftrightarrow x_j  
\end{equation}
where $\hat{\mathcal{X}} = \mathcal{X} \backslash \{ x_i, x_j \}$, i.e, the difference set resulting from subtracting $\{ x_i, x_j \}$ from $\mathcal{X}$, and $x_i \leftrightarrow x_j$ means that there is a causal edge of unknown direction between $x_i$ and $x_j$.
A conditional independence test can be performed for the null hypothesis that the correlation is equal to zero.
\end{definition}

\subsection{Structural Causal Models}\label{sec:Preliminary Structural Causal Model}
\begin{definition}[Structural Causal Model]
A structural causal model (SCM) is defined as 
\begin{equation}
       \mathcal{M}:= \left\langle \mathcal{X}, \mathcal{U}, \mathbb{P}_{\mathcal{U}}, \mathcal{F} \right\rangle,
\end{equation}
where 
\begin{enumerate}
\item $\mathcal{X}$ is a set of observed variables;
\item $\mathcal{U}$ is a set of unobserved exogenous noise;
\item $\mathbb{P}_{\mathcal{U}}$ is the distribution for $\mathcal{U}$;
\item $\mathcal{F}$ is $\mathcal{X} \times \mathcal{U} \rightarrow \mathcal{X}$ is a measurable function that specifies the causal mechanism.
\end{enumerate}
\end{definition}

\begin{definition}[Structural equations]\label{def of Structural equations}
Let $\mathcal{M}$ be an SCM, the structural equations of the $\mathcal{M}$ are the set of equations
\begin{equation}\label{structural equation}
    x_i = f_i(\boldsymbol{x_{pa_i}}, u_i), x_i \in \mathcal{X}, u_i \in \mathcal{U},
\end{equation}
where $\boldsymbol{x_{pa_i}}$ is the set of variables that have causal effect on $x_i$, i.e., the parent of $x_i$. The $u_i$ is the unobserved exogenous noise associated with $x_i$.
\end{definition}

TCG suffers from the limitation of only expressing the presence or absence of causal relationships, without accurately representing the magnitude and specific functions of causal effects. On the other hand, SCM can express specific causal relationships using various functions however the validity of the model is constrained by the rationality of the human-defined edges. 

In this paper,  our TVCG uses a Hybrid Causal Graph design, which merges TCG with SCM techniques. This hybrid model incorporates TCG's causal discovery mechanism and enhances it with SCM’s soft  (deterministic or probabilistic) edge-fitting methods, thus leveraging the strengths of both methods.

\section{The CLIS Method: TVCG} \label{sec:method} 
This section illustrates the proposed tabular-visual causal graph (TVCG) for the Causal Longitudinal Image Synthesis (CLIS) task. TVCG focuses on constructing a temporal causal graph model, which encompasses both tabular data and MRI, with the medical image serving as the target output. As shown in Fig.~\ref{fig:overview}, there are two training processes:\textbf{ (1) tabular-only causal graph (TOCG) establishment} and \textbf{(2) tabular-visual causal graph (TVCG) establishment.} In this section, we will first introduce the data we use, then detail the two training processes, and finally, describe TVCG's intervened inference process for CLIS, and finally a downstream task to apply the prediction results of TVCG.

\subsection{Data}
To provide a clearer understanding of our TVCG, we use Alzheimer's Disease (AD) as a working example, utilizing the widely used AD dataset, ADNI~\cite{ADNI}. Balancing variable coverage and model complexity, we select 11 variables from the ADNI~\cite{ADNI} dataset for our experiment, denoted as $d=11$. For a better understanding, we list all these 11 variables in Table~\ref{tbl:variable}, and $x_d^t$ represents the $d$-th observed variable at time $t$ for an AD patient. Now, the union of them, $\boldsymbol{x}$, is a $d$-dimension variable: 
\begin{equation}
    \boldsymbol{x}^t=[x_1^t, ... , x_d^t]=[\mathcal{S}^t, \mathcal{B}^t,\mathcal{V}^t,\mathcal{X}^t].
\end{equation}
Note that $x_i$ can represent different data types; therefore, we further divide them into four distinct groups: $\mathcal{S}$ for demographic variables, $\mathcal{B}$ for bio-marker, $\mathcal{V}$ for brain and substructure volumes, and $\mathcal{X}$ for MRI. The total intracranial volume (TIV), ventricle volume (VV) and grey matter volume (GMV) are chosen as they are closely related to the AD progression~\cite{GM1, GM2, GM3, ventricles1, ventricles3}.
\begin{table}[t]
\caption{The definition of variables $x_i$.}\label{tbl:variable}
\begin{tabular*}{0.49\textwidth}{lll}
\toprule%
variable & definition & unit \\ 
\midrule
\multicolumn{3}{l}{$\mathbf{demographic \ variables} \ \mathcal{S}$}  \\
$x_{1}^t$ & age at time t& years\\
$x_{2}^t$ & biological gender: 0 for male and 1 for female & -\\
$x_{3}^t$ & the duration of receiving education & years\\
$x_{4}^t$ & the apolipoprotein E (APOE) gene (0/1/2) & - \\
\midrule
\multicolumn{3}{l}{$\mathbf{biomarker \ level}  \ \mathcal{B}$}  \\
$x_5^t$ & the $A\beta_{1-42}$ level in cerebro-spinal fluid & pg/ml \\
$x_6^t$ & the $\tau$ protein level in cerebro-spinal fluid & pg/ml \\
$x_7^t$ & the $p\tau_{181}$ protein level in cerebro-spinal fluid & pg/ml \\
\midrule
\multicolumn{3}{l}{$\mathbf{brain \ and \ substructrure \ volume} \ \mathcal{V}$}  \\
$x_8^t$ & the total intracranial volume (TIV) & ml \\
$x_9^t$ & the ventricle volume (VV) & ml \\
$x_{10}^t$ & the grey matter volume (GMV) & ml \\
\midrule
\multicolumn{3}{l}{$\mathbf{T_1\text{-}weighted \ MR \ image \ data} \ \mathcal{X}$}  \\
$x_{11}^t$ & MR image cropped to $192\times192\times224$ \\ 
 & with every voxel normalised to [0, 1] & mm$^3$\\
\bottomrule
\end{tabular*}
\end{table}

\subsection{Tabular-Only Causal Graph (TOCG) Establishment}~\label{sec:nonimage} 
Given the significant gap between tabular and image variables, we opt to first establish a TOCG $\mathscr{G} [\mathcal{S}, \mathcal{B}, \mathcal{V}]$  instead of directly building the entire TVCG $\mathscr{G} [\mathcal{S}, \mathcal{B}, \mathcal{V}, \mathcal{X}]$  for our CLIS. Specifically, there are two steps of the TOCG establishment: causal edge discovery, and structural causal model fitting, with the aid of pre-defined assumptions.

\subsubsection{Pre-defining Assumptions}\label{sec:Assumptions predefined}
Besides the Causal Sufficiency and Faithfulness assumptions defined in section~\ref{sec:Preliminary Window Causal Graph}, we further make the following three more assumptions based on the AD scenario.

 \begin{assumption}[Markov Property]\label{Markov Assumption}
The probability distribution of future states of $\boldsymbol{x}$ depends only upon the present state, i.e., 
\begin{equation}
    {\forall} \boldsymbol{x}, P\big(\boldsymbol{x}^t|\boldsymbol{x}^{t-1}, ..., \boldsymbol{x}^{1}\big) =  P\big( \boldsymbol{x}^t|\boldsymbol{x}^{t-1}\big),
\end{equation}
which implies that there is no direct link when the time lag $\tau$ is greater than 1.
\end{assumption}

\begin{assumption}[Self Temporal Causality]\label{Self Causality}
There always exists a causal effect from $\boldsymbol{x}_i^{t-1}$ on $\boldsymbol{x}_i^{t}$. i.e., 
\begin{equation}
    \forall \boldsymbol{x}_i, \forall t,  \boldsymbol{x}_i^{t-1} \xrightarrow{e} \boldsymbol{x}_i^{t},~~e\in \boldsymbol{E},
\end{equation}
where $\boldsymbol{E}$ is the edge set of the causal graph and the only `exception' is that for a constant factor, such as gender, there is no self-temporal causality.
\end{assumption}

\begin{assumption}[Prior Causal Hierarchy]\label{PK}
There is a causal hierarchy among different groups of variables, inspired from the prior knowledge~\cite{event_order1, event_order2, event_order3}. Specifically,

(i) The demographic variables $S$ (e.g., gender) occupy the highest level: the edges from other modalities' variables to demographic variables are prohibited:
\begin{equation}
    \forall \boldsymbol{x}_i \not\in \mathcal{S}, \ \forall \boldsymbol{x}_j \in \mathcal{S}, \boldsymbol{x}_i \xrightarrow{\hat{e}} \boldsymbol{x}_j, ~~ \hat{e} \notin \boldsymbol{E}.
\end{equation}

(ii) The biomarker variables $B$ (e.g., protein percentage) are with the second highest level: the edges from other modalities' variables (except demographic) to biomarker variables are prohibited. i.e., 
\begin{equation}
    \forall \boldsymbol{x}_i \not\in (\mathcal{S}\cup B), \ \forall \boldsymbol{x}_j \in \mathcal{B}, \boldsymbol{x}_i \xrightarrow{\hat{e}} \boldsymbol{x}_j,~~ \hat{e} \notin \boldsymbol{E}.
\end{equation}
\end{assumption}

\begin{figure*}[t]
  \centering\includegraphics[width=0.95\textwidth]{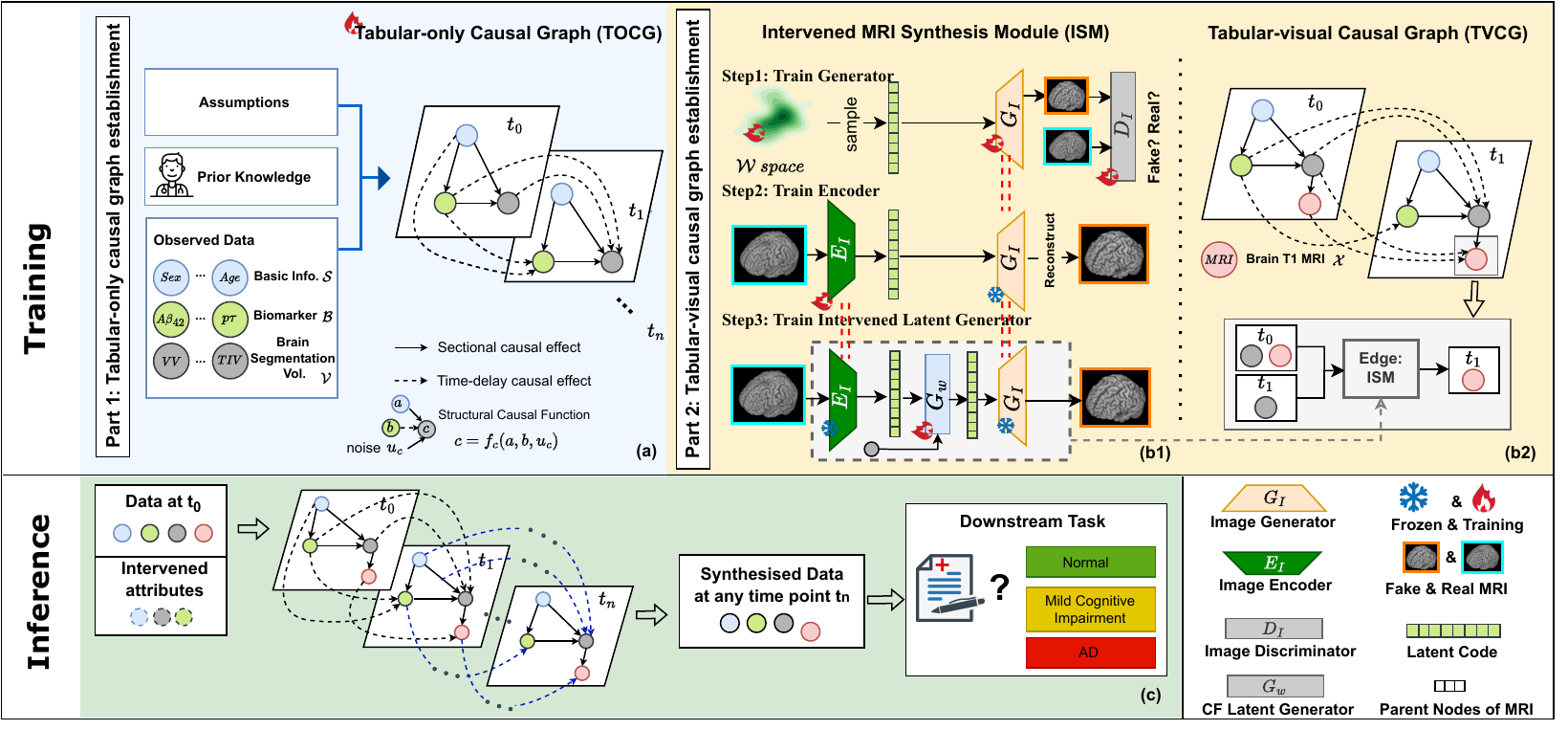}
      \caption{The training and inference phases of the proposed tabular-visual causal graph (TVCG). The training includes two parts. (1) The tabular-only causal graph construction. Firstly causal discovery algorithms are applied to the observed tabular variables with assumptions and prior medical knowledge. Then for the recovered edges, the generative function is fitted. (2) The tabular-visual causal graph (TOCG) construction. An Intervened MRI Synthesis Module (ISM) is trained for the MRI generative. ISM is severed as an edge between tabular volume variables and MRI thus building TVCG over TOCG. The ISM training is a 3-step process:  training an image generator, an image encoder, and an intervened latent generator. During inference, a set of baseline and optional intervention variables are input into the trained TVCG. The model computes the intervened tabular variables using the TOCG within TVCG, and the ISM generates the intervened MRI. These synthesized results are then applicable to downstream classification tasks.}
  \label{fig:overview}
\end{figure*}


\subsubsection{Causal Edge Discovery}\label{sec:causal discovery method}
The causal edge discovery is used to model the presence of causal relationships as edges $E$ in a causal graph.
Specifically, by applying the Markov Property (Assumption~\ref{Markov Assumption}), the causal discovery problem can be simplified to focus on the discovery of causal relationships with a maximum time lag of $\tau_{max} = 1$. Thus, for an observed trajectory $\boldsymbol{x}^t$ with a time series length of $T$, the observational data can be arranged into pairs:
\begin{equation}
    \boldsymbol{x}_{split} = \{(\boldsymbol{x}^1, \boldsymbol{x}^2), ... , (\boldsymbol{x}^{T-1}, \boldsymbol{x}^{T})\}, 
\end{equation}
where each $(\boldsymbol{x}^{t-1}, \boldsymbol{x}^{t})$ is an observational data pair from two consequence time steps ($t-1$ and $t$).  
Thus the causality discovery can be performed with the help of Assumptions~\ref{Self Causality} and ~\ref{PK}.

Technically, we employ three widely used algorithms: FCI~\cite{FCI} (constraint-based), GES~\cite{GES} (score-based), and DirectLiNGAM~\cite{DirectLiNGAM} (functional-causal-models-based). The results from these algorithms are integrated via a voting mechanism.
The implementation leverages the Causal-learn package~\footnote{https://github.com/py-why/causal-learn}, with several modifications made to the algorithms to comply with the constraints outlined in Assumptions~\ref{Self Causality} and~\ref{PK}. Detailed information on the algorithms is available in the code, and the validity of the learned causal model is discussed in the Experiments Section. The resulting causal graph is presented in Figure ~\ref{causal graph}.

For testing conditional independence in the partial correlation analysis, as mentioned in Section~\ref{sec:Preliminary Causal Discovery}, a $t$-test is utilized. We set the p-value threshold at 0.05 to reject the null hypothesis.



\subsubsection{Structural Causal Model Fitting}\label{sec:Structural causal model fitting}
After modeling the presence or absence of causal relationships (edges) via causal edge discovery, we further model these relationships (edges $E$) via deterministic or probabilistic functions $f$ as in SCM.
In accordance with the TCG, the traditional SCM as Equ.~\eqref{structural equation} is extended to introduce time modeling interval $\delta t$ for continuous-time as 

\begin{equation}\label{temporal structural equation}
    x_i^{t_1} = f_i(\boldsymbol{x_{pa_i}}, u_i, \delta t), x_i \in \mathcal{X}, u_i \in \mathcal{U},
\end{equation}
where $\boldsymbol{x_{pa_i}}$ is the set of variables that have causal effect on $x_i^{t}$.  $x_i^{t_1}$ is the variable $i$ at time $t_1$. $t_0$ is the time point of previous session and $\delta t = t_1 - t_0$.

We model the function $f_i$ as a form of (1) Linear function, (2) Multi-layer Perceptron (MLP), or (3) Normalizing flow. Besides, we compare with Recurrent Neural Network (RNN)~\cite{RNN} and Long short-term memory (LSTM)~\cite{LSTM}, two bi-directional artificial neural networks that use an internal state to process arbitrary sequences. The details of the exact model of each function is provided in the supplementary. 

Here we take MLP $\hat{f}_{mlp}$ as an example. We assume an additive noise model~\cite{ADM} as
\begin{equation}
   x_i^{t_1} = f_i(\boldsymbol{x_{pa_i}}, \delta t, u_i) := \hat{f}_{mlp}(\boldsymbol{x_{pa_i}}, \delta t) + u_i.
\end{equation}
The \textbf{training objective function} is with $L_2$ loss as 
\begin{equation}
    \mathcal{L} = \mathcal{L}_{2}(x_i^{t_1}, \hat{f}_{mlp}(\boldsymbol{x_{pa_i}}, \delta t)),
\end{equation}
where $x_i^{t_1}$ is the observation data.


\subsection{Tabular-Visual Causal Graph (TVCG) Establishment} \label{sec:image}
In this paper, we aim to process a variety of variables, including both tabular data $\{\mathcal{S}, \mathcal{B}, \mathcal{V}\}$ and MRI $\{\mathcal{X}\}$. However, a TOCG 
is insufficient for this purpose, we still need to build a TVCG. 
A straightforward way to achieve this would be to treat MRI as a variable similar to the tabular variables. This presents a significant challenge due to the giant dimensional gap between tabular variables and MRI. In this paper, we propose to use an intervened MRI synthesis module (ISM) to bridge the gap.

Specifically,  we first select three tabular variables that are directly associated with MRI from the established tabular-only causal graph. These variables are Total Intracranial Volume (TIV) $x_{8}^t$, Ventricular Volume (VV) $x_{9}^t$, Gray Matter Volume (GMV) $x_{10}^t$. Then our ISM modifies the MRI in the former time steps to synthesize a new MRI in the current time step according to the three variables. In this way, the ISM serves as an edge between the tabular-only causal graph and MRI data $x_{11}^t$,  thereby realizing the TVCG. 

In the following, we will first introduce the training process of ISM and then detail how to get the final TVCG.


\subsubsection{Intervened MRI Synthesis Module (ISM)}
As shown in Fig.~\ref{fig:overview}, We train our ISM in three steps: (1) Train a latent-to-image generator that can map a latent space to image space, (2) Train an image-to-latent encoder to perform an inverse mapping from image space back to a latent space, and (3) Train a volume-variable-to-latent generator to generate latent code based on volume variables.  After that, We can employ the trained latent code generator to transform any volume variables $\mathcal{V}=\{x_{8}^t,x_{9}^t,x_{10}^t\}$ into latent codes and then synthesize images based on the latent codes via the trained generator and encoder. When the volume variables are derived from our causal graph, the synthesized image becomes a \textbf{causally-synthesized} image. In the following, we will introduce each step.




\underline{Step1}: Train the latent-to-image generator.\label{styleGAN and generator}
Unlike typical natural image synthesis tasks, MRI data for training are often limited, making it challenging to directly train an image-latent-image model. To mitigate this issue, we train our generator and encoder separately. In this way, the latent-to-image generator can be effectively trained by sampling latent codes from a distribution, thereby overcoming the constraints of limited data availability.

A style-based generative adversarial network (StyleGAN) is trained as the latent-to-image generator for its astonishing quality of generated images~\cite{karras2019style, Controllable_Image_Synthesis}.
The StyleGAN generator $f_{style}$ firstly maps a multi-dimensional Gaussian distribution $\mathcal{Z}$ to the latent space $\mathcal{W}$ by a mapping network $f_{map}$:
\begin{equation}
    f_{map}: \mathcal{Z} \rightarrow \mathcal{W}, \mathcal{Z} \subseteq \mathbbm{R}^{128}, \mathcal{W} \subseteq \mathbbm{R}^{128}, 
\end{equation}
and then a series of styled convolution blocks generate images $\mathbf{I}$ from the sampled latent variable $\boldsymbol{w}$ with Gaussian noise $\boldsymbol{n}$ added:
\begin{equation}
    \mathbf{I} = \mathbf{G_I}(\boldsymbol{w}, \boldsymbol{n}), \mathbf{I} \in \mathbbm{R}^{192\times224\times192}, \boldsymbol{w} \sim \mathcal{W},
\end{equation}
where $\mathbf{G_I}$ represents the styled convolution blocks. 

A CNN discriminator $\mathbf{D_I}$  evaluates the synthesized image distribution and try to distinguish it from the true data distribution. The generator and discriminator are trained simultaneously but independently in an adversary way.

The \textbf{training objective functions} of discriminator $\mathbf{D_I}$ and generator $\mathbf{G_I}$ are given by
\begin{equation}\label{d loss}
\begin{split}
\mathcal{L}(\mathbf{D_I}) = & \  \mathbb{E}_{\boldsymbol{w}\sim\mathcal{W}}[\mathbf{D_I}(\mathbf{G_I}(\boldsymbol{w}, \boldsymbol{n}))] -  \mathbb{E}_{\mathbf{I} \sim data}[\mathbf{D_I}(\mathbf{I})] \\  & + \lambda_{grad}\mathbf{E}_{\tilde{\mathbf{I}}}[(\left \| (\nabla \mathbf{D_I})_{\tilde{\mathbf{I}}} \right \| - 1)^2],
\end{split}
\end{equation}
\begin{equation}\label{g loss}
\mathcal{L}(\mathbf{G_I}) = -\mathbb{E}_{\boldsymbol{w}\sim\mathcal{W}}[\mathbf{D_I}(\mathbf{G_I}(\boldsymbol{w}, \boldsymbol{n}))],
\end{equation}
where the first and second terms in Equ.~\eqref{d loss} represent the discriminator output expectation of generated and real distributions respectively. The third term is the Gradient Penalty. 

\underline{Step2}: Train the image-to-latent encoder.\label{sec:styleGAN's encoder}
After we get a well-trained latent-to-image generator based on the Gaussian distribution, we now train an image-to-latent encoder to map the MRI to the latent code. In this way, we can modify the MRI (or intervene) by introducing the three variables into the latent code and synthesis the new MRI (or intervened MRI). 

Specifically, with the well-trained StyleGAN as a fixed generator, an encoder $\mathbf{E_w}$ is trained to project MRI $\mathbf{I}$ into the latent space $\hat{\boldsymbol{w}}$: 
\begin{equation}
    \hat{\boldsymbol{w}} = \mathbf{E_w}(\mathbf{I}), \hat{\boldsymbol{w}} \in \mathcal{W}^* \subseteq \mathbf{R}^{12\times128}
\end{equation}
where $\hat{\boldsymbol{w}}$ is the mapping latent code, laying in the $\mathcal{W}*$ space, which should align with the latent code space $\mathcal{W}$ in the fixed styleGAN. It is worth noting that the latent code $\boldsymbol{w} \in \mathcal{W} \subseteq \mathbb{R}^{128} $ only have 128 dimension, they are duplicated $12$ times to fit the input the generator, while $\mathcal{W}^* \subseteq \mathbb{R}^{12\times128}$ owns more expressiveness, i.e., the cardinality of set $\mathcal{W}$ is greater than $\mathcal{W}^*$:$|\mathcal{W}| < |\mathcal{W}^*|.$
Tov et al's work\cite{e4e} and supplementary material provide more information about space $\mathcal{W}^*$. 

Our encoder is based on a backbone with a feature pyramid that generates three levels of feature maps~\cite{pSp}. The details of the encoder can be found in the supplementary material.

The \textbf{training objective functions} of image encoder $\mathbf{E_w}$ and latent code discriminator $\mathbf{D_w}$ are given by
\begin{equation}\label{Ew loss}
\begin{split}
    \mathcal{L}(\mathbf{E_w}) = & \ \mathcal{L}_1(\mathbf{G}(\mathbf{E_w(\mathbf{I})}, \boldsymbol{n}), \mathbf{I}) \\ &+ 
    \lambda_{freq}\mathcal{L}_1(\mathscr{F}(\mathbf{G}(\mathbf{E_w}(\mathbf{I}), \boldsymbol{n})), \mathscr{F}(\mathbf{I})) \\
      &- \lambda_{ad}\mathbb{E}_{\mathbf{I} \sim data}[\mathbf{D_w}(\mathbf{E_w}(\mathbf{I}))] \\ &+ \lambda_{reg}\mathcal{L}_{reg}(\mathbf{E_w(\mathbf{I})}),
\end{split}
\end{equation}
\begin{equation}\label{Dw loss}
\begin{split}
    \mathcal{L}(\mathbf{D_w}) = & \ \mathbb{E}_{\mathbf{I} \sim data}[\mathbf{D_w}(\mathbf{E_w}(\mathbf{I}))] \ - \ \mathbb{E}_{\boldsymbol{w} \sim \mathcal{W}}[\mathbf{D_w}(\boldsymbol{w}))] \\ &+ \lambda_{grad} \mathbf{E}_{\mathcal{W}}[(\left \| (\nabla \mathbf{D_w})_{\boldsymbol{w}} \right \| - 1)^2].
\end{split}
\end{equation}
where the L1 loss in both the image domain and frequency domain ($\mathcal{F}$: Fourier Transform) enables learning on a pixel-wise basis while also avoiding blurry images that result from the loss of important frequencies. A regularisation term and a latent discriminator trained adversarially is applied to restrict $\mathcal{W}^*$ not to deviate from $\mathcal{W}$ too far. 

For an MRI, the trained encoder projects it to the latent space, and the Gaussian noise added to the image generator is obtained by optimizing 
\begin{equation}
    \boldsymbol{\hat{n}} = \arg \min_{\boldsymbol{n}}{\mathcal{L}_1} (\mathbf{I}, \boldsymbol{G_I}(\boldsymbol{E_w}(\mathbf{I}), \boldsymbol{n})).
\end{equation}

\underline{Step3}: Train the intervened latent generator.
As mentioned in step 2, we want to introduce the three variables $V$ into the latent code to synthesize the new MRI (or intervened MRI).  However, the variables have different shapes from our latent code. Hence, a V-to-latent generator is trained to generate latent variables based on V, i.e., {\it brain and substructure volumes}. We term it as an Intervened latent generator since the three variables $V$ are a kind of intervention to modify the MRI in the former timestep to get a new MRI.

Specifically, we use a lightweight 5-layer MLP as the Intervened latent generator. This design avoids the overfitting issue for the low-dimensional latent space. If a patient has undergone MRI at $T$ different time steps $T=[t_1, ..., t_T]$, the training objective function of the Intervened latent code generator $\mathbf{G_w}$ is given by:
\begin{equation}
\begin{aligned}
&\mathcal{L}(\mathbf{G_w})=\sum_{t_n \in T} \sum_{\substack{t_m \in T }} \mathcal{L}_1 \left(\boldsymbol{\hat{w}}^{t_n},\boldsymbol{w}^{t_n}(t_m)\right),~t_m < t_n \\
&\boldsymbol{w}^{t_n}(t_m)=\mathbf{G_w}(\boldsymbol{\hat{w}}^{t_m}, \boldsymbol{V}^{t_m}, \boldsymbol{V}^{t_n}),\\
&\boldsymbol{\hat{w}}^{t_m}=E_w(\mathbf{I}^{t_m}),\boldsymbol{\hat{w}}^{t_n}=E_w(\mathbf{I}^{t_n}),
 \end{aligned}
\end{equation}
where $\hat{\boldsymbol{w}}^{t_m}$ and $\hat{\boldsymbol{w}}^{t_n}$ denotes the mapping latent code of the MRI at time points $t_m$ and $t_n$, respectively, projected by the trained encoder $\mathbf{E_w}$. $\boldsymbol{w}^{t_n}(t_m)$ is the output of $\mathbf{G_w}$. We train $\mathbf{G_w}$ 
to approximate $\boldsymbol{w}^{t_n}(t_m)$ to the GT $\hat{\boldsymbol{w}}^{t_n}$.

Likewise, $\boldsymbol{V}^{t_n} = \{x_{8}^{t_n},x_{9}^{t_n},x_{10}^{t_n}\}$ and  $\boldsymbol{V}^{t_m} = \{x_{8}^{t_m},x_{9}^{t_m},x_{10}^{t_m}\}$ denotes the corresponding segmented brain volumes at visit time point $t_n$ and $t_m$. In our case, $t_m \textless t_n$, which means $\mathbf{G_w}$ are trained to transfer the former timestep MRI's latent code $\hat{\boldsymbol{w}}^{t_m}$ to the latter timestep MRI's latent code $\hat{\boldsymbol{w}}^{t_n}$ conditioned in the modification $\boldsymbol{V}_j^{t_m} \rightarrow \boldsymbol{V}_j^{t_n}$.

During training, for a patient with $T$ scans in total, we randomly select two scans $t_n$ and $t_n$ in each training epoch. 

\subsubsection{TVCG Based on ISM}
After we trained the ISM, we can directly add the MRI into the TOCG by treating ISM as an edge between the three volume variables and MRI.

Specifically, the three tabular volume variables $V$: TIV($x_8$), VV($x_9$), GMV($x_{10}$) serve as the parent nodes (higher level variables) of the MRI $X=\{I\}=\{x_{11}\}$, the edges between them exactly define the trained ISM.
\begin{equation}
\begin{aligned}
    &\mathscr{G}[\mathcal{S}, \mathcal{B}, \mathcal{V}]+ \mathcal{X} \xrightarrow{e=ISM} \mathscr{G} [\mathcal{S}, \mathcal{B}, \mathcal{V}, \mathcal{X}], ISM \in E\\
    &\mathcal{X}^{t+1}=\{x_{11}^{t+1}\}    =ISM(\mathcal{V}^t, \mathcal{V}^{t+1}, \mathcal{X}^t),\\
    &\mathcal{V}^t=\{x_8^t,x_9^t,x_{10}^t\},X^t=\{\mathbf{I}^t\}=\{x_{11}^t\}.
\end{aligned}
\end{equation}

Now we obtain the TVCG $\mathscr{G} [\mathcal{S}, \mathcal{B}, \mathcal{V}, \mathcal{X}]$ and any modification over the node of the graph (e.g, over $\mathcal{S}$ and $\mathcal{B}$) causally affects the $\mathcal{V}$ and further influences $\mathcal{X}$. The influence in $\mathcal{X}$ caused by the graph is termed as an intervention in our work.

\subsection{Intervened Inference Processes}
With the TVCG, we can address various counterfactual 'if' scenarios. These scenarios are categorized into two types based on their underlying hypotheses: 1) Questions related to MRI based on the brain volume hypothesis, and 2) Questions based on other hypotheses. For the first type, we directly synthesize the Intervened MRI using the ISM. For other hypothesis-based questions, the approach varies: if they pertain to tabular variables, we find answers using our TOCG; if they relate to MRI, we first estimate the corresponding brain volume via TOCG and then synthesis the Intervened MRI using ISM.
In this part, we introduce the inference process of ISM and TVCG, respectively. We set the time interval as 1 during inference.

\subsubsection{ISM Inference for Brain Volume Hypothesis}
In ISM, any intervention (or hypothesis) on volume variables $\boldsymbol{V}^{t_m} \rightarrow \boldsymbol{V}^{t_n}$ results in a corresponding intervention in the MRI $\mathbf{I}^{t_m}\rightarrow \mathbf{I}^{t_n}$, 
\begin{equation}
    \mathcal{X}^{t_n}=\{\mathbf{I}_{ism}^{t_n}\} = \mathbf{G_I}(\mathbf{G_w}(\mathbf{E_w}(\mathbf{I}^{t_m}), \boldsymbol{V}^{t_m},
    \boldsymbol{V}^{t_n}), \hat{n}),
\end{equation}
where $\mathbf{I}^{t_m}$ is the MRI at $t_m$ and $\mathbf{I}_{ism}^{t_n}$ is the synthesized MRI at $t_n$. In fact, ISM focuses on the difference between $\boldsymbol{V}^{t_m}$ and $\boldsymbol{V}^{t_n}$, and synthesises $\mathbf{I}^{t_n}_{ism}$ on the base of $\mathbf{I}^{t_m}$.

\subsubsection{TVCG Inference for Other Hypotheses}
In our TVCG, any intervention on $\mathcal{S}$- or $\mathcal{B}$-variable causally leads to the change of other $\mathcal{V}$ variables: $\boldsymbol{V}^{t_m} \rightarrow \boldsymbol{V}^{t_n}$.
The $\mathcal{V}^{t_n}$ are predicted first based on TOCG and then MRI $\mathcal{X}^{t+1}$ is synthesized via ISM.
\begin{equation}
    \{\mathcal{S}^{t_n}, \mathcal{B}^{t_n}, \mathcal{V}^{t_n}\} = \mathcal{G}[\mathcal{S}^{t_m}, \mathcal{B}^{t_m}, \mathcal{V}^{t_m}],
\end{equation}
\begin{equation}\label{equ: synthesized image}
\mathcal{X}^{t_n}=\{\mathbf{I}_{ism}^{t_n}\}=ISM(\mathcal{V}^t_m, \mathcal{V}^{t_n}, \mathcal{X}^t_m).
\end{equation}

Besides the longitudinal MRI counterfactual synthesis, TVCG can also answer the counterfactual question for tabular variables. For example, for a patient
with AD, how will be the brain grey
matter atrophy in a year if intervened on the A-beta level in
cerebrospinal fluid? The result is computed as
\begin{equation}
\begin{split}
    a \rightarrow x_{5}^{t_m},~~&x_5 \in \mathcal{B}^{t_m} \\
    \{\mathcal{S}^{t_n}, \mathcal{B}^{t_n}, \mathcal{V}^{t_n}\} &= \mathcal{G}[\mathcal{S}^{t_m}, \mathcal{B}^{t_m}, \mathcal{V}^{t_m}](x_{5}^{t_m}), \\
    \Delta x_{10} = x_{10}^{t_n} - x_{10}^{t_m},&~~ x_{10}^{t_m} \in \mathcal{V}^{t_m},~~x_{10}^{t_n} \in \mathcal{V}^{t_n}, \\
\end{split}
\end{equation}
where A-beta level$(x_5^t$) is set to $a$ and $\Delta x_{10}$ is the brain grey
matter atrophy amount.

\subsection{Downstream Tasks}
To demonstrate that the generated images possess clinical relevance in predicting the future progression of a patient's condition, we employ the generated data to predict the patient's diagnostic ${y}^t$ classification as downstream tasks. 
\begin{equation}
    {y}^t=\left\{
    \begin{array}{rcl}
    0&& \text{Normal \ Control};\\
    1&& \text{Mild \ Cognitive \ Impairment};\\
    2&& \text{AD}.
    \end{array} \right.
\end{equation}
\begin{equation}
\begin{split}
    \hat{y}^t = F(\phi(\mathcal{X}^{t}), [\mathcal{S}^{t}, \mathcal{B}^{t}, \mathcal{V}^{t}]),
\end{split}
\end{equation}
where $\hat{y}^t$ is the predicted  diagnosis class, $\phi$ is a neural network ({\it e.g.}, DenseNet121) to extract the image feature, and $F$ is a classification network which takes in the extracted image feature and tabular variables $x_j^t \in \{\mathcal{S}, \mathcal{B}, \mathcal{V}\}$. 
We train the $\phi$ and $F$ jointly on real data. 
\begin{equation}
    \mathcal{L}(\phi, F) = l_{BCE}(y^t, F(\phi(\mathcal{X}^{t}), [\mathcal{S}^{t}, \mathcal{B}^{t}, \mathcal{V}^{t}])).
\end{equation}

\section{Experiments} \label{sec:expe}
\subsection{Datasets and Training Details}
We train our models using the ADNI dataset (adni.loni.usc.edu), and to assess model generalization, we test them on two independent datasets: OASIS3 and NACC. Detailed information about these publicly available datasets is presented in Table~\ref{Dataset info}. The ADNI dataset is our primary choice for model construction as it encompasses all variables listed in Table~\ref{tbl:variable}. The NACC dataset, lacking longitudinal data, and the OASIS dataset, missing biomarker variables and primarily comprising healthy individuals, is used for limited testing purposes. OASIS is specifically employed for evaluating the image-related components of our model, not for testing causality and downstream tasks.
For imaging processing, all T$_1$ images are skull-stripped using ROBEX~\cite{robex}, aligned to the MNI152 space, resampled to 1mm isotropic resolution using ANTs~\cite{ANTs}, cropped to dimensions of 192×224×192, and normalized in voxel values to the range [0, 1]. We segment the processed MRI to obtain grey matter volume with ANTs and ventricle volume through a CNN segmentation model~\cite{seg_ventricle}.

\begin{table}[t]
\caption{The Basic Information of Three Datasets}\label{Dataset info}
\begin{tabular*}{0.47\textwidth}{l l l l}
\toprule
\multicolumn{4}{c}{\textbf{ADNI}} \\
Category & NC & MCI & AD \\
\# of subjects & 739 & 981 & 383 \\
\# of sessions & 2675 & 3388 & 1570 \\
Sex (Female/Male) & 327/412 & 580/401 & 219/164 \\
APOE (0/1/2) & 516/202/21 & 493/379/109 & 123/181/79 \\
Age at baseline & 73.14(±6.23) & 72.77(±7.54) & 74.83(±7.90) \\
Years of education & 16.54(±2.56) & 15.99(±2.74) & 15.28(±2.91) \\
\midrule
\multicolumn{4}{c}{\textbf{NACC}} \\
Category & NC & MCI & AD \\
\# of subjects & 1682 & 484 & 667 \\
\# of sessions & 3463 & 894 & 1179 \\
Sex (Female/Male) & 1030/652 & 228/256 & 297/370 \\
APOE (0/1/2) & 1047/540/50 & 251/166/44 & 257/264/94 \\
Age at baseline & 68.85(±10.15) & 72.14(±9.01) & 71.67(±9.34) \\
Years of education & 16.16(±4.01) & 15.92(±6.32) & 15.55(±7.31) \\
\midrule
\multicolumn{4}{c}{\textbf{OASIS}} \\
Category & NC & MCI & AD \\
\# of subjects & 495 & 24 & 62 \\
\# of sessions & 1267 & 61 & 161 \\
Sex (Female/Male) & 295/200 & 11/13 & 19/43 \\
APOE (0/1/2) & 252/129/20 & 12/10/0 & 25/26/5 \\
Age at baseline & 67.74(±8.67) & 70.22(±7.76) & 73.22(±6.85) \\
Years of education & 16.08(±2.55) & 14.71(±2.47) & 15.79(±2.51) \\
\bottomrule
\end{tabular*}
\end{table}

For the learning of MLP, NF, LSTM and RNN in Section~\ref{sec:Structural causal model fitting}, Adam is used with a learning rate of 0.001 and $(\beta_1, \beta_2) = (0.9, 0.99)$. All models are trained for 5,000 epochs.

In Section~\ref{sec:styleGAN's encoder}, we detail the progressive training approach for StyleGAN, starting with low-resolution images and progressively doubling the resolution at each phase. Beginning at $6\times7\times6$, the process culminates in generating images of $192\times224\times192$ at a 1mm isotropic resolution over six phases. Due to increasing memory demands, the minibatch size is adjusted through the phases as ${ 2048, 2048, 1024, 256, 64, 24 }$. Both the discriminator and generator are optimized using Adam with $(\beta_1, \beta_2) = (0, 0.99)$. Learning rates are set at 0.001 for phases 1-4, 0.015 for phase 5, and 0.002 for the final phase. Each phase involves training on 2,000,000 samples, except the last one, which uses 150,000 mini-batches. The $\lambda_{grad}$ parameter in the discriminator's loss equation (Equ.~\eqref{d loss}) is fixed at 10.

The image encoder $\mathbf{E_w}$ and latent code discriminator $\mathbf{D_w}$ are trained independently but simultaneously, as in StyleGAN.  For the encoder loss, Equ.~\eqref{Ew loss}, the $\lambda_{ad}$ is $0.005$, $\lambda_{freq}$ is $0.5$ and $\lambda_{reg}$ is $0.0001$. The $\lambda_{grad}$ in w discriminator loss, Equ.~\eqref{d loss}, is 10. The Optimizer Adam is used with a learning rate of 0.00001 for $\mathbf{E_w}$ and 0.00002 for $\mathbf{D_w}$ and  $(\beta_1, \beta_2) = (0.9, 0.99)$ for both. The mini-bath size is 16 and the training comprises the iterations of 100,000 batches. The code for this study is publicly available\footnote{https://github.com/jessyblues/Causal-Longitudinal-Image-Synthesis.git}.

\subsection{Results}
\subsubsection{Causal Graph and Structural Causal Model}\label{sec:results causal graph}
\underline{Causal edge validity}:
As outlined in Section~\ref{sec:causal discovery method}, we utilize multiple causal discovery methods on the ADNI dataset to verify causality relationships, which are then tested on the NACC dataset. For causal discovery, the ADNI data is split into adjacent session pairs.

In practical terms, the ADNI dataset is divided into a 4:1 ratio for training and validation sets. We apply three causal discovery methods, FCI, GES, and DirectLiNGAM, to the training set. Due to the absence of longitudinal data in NACC, we only test sectional causal relationships (causal edges with time lag 0) in this dataset.
\begin{table}[t]
\caption{The training and test sets for causal discovery.}\label{Dataset split}{
\begin{tabular*}{0.47\textwidth}{l l r r r}
\toprule%
 & & \# Subject & \# Data Pairs & Interval Years \\
 Training & ADNI & 364 & 560 & 2.03$_{\pm1.21}$ \\
 Test & ADNI & 91 & 147 & 1.95$_{\pm 1.19}$\\
 Test & NACC & 97 & 97 & - \\
\bottomrule
\end{tabular*}}
\end{table}

\begin{figure}[t]
  \centering
  \includegraphics[width=0.4\textwidth]{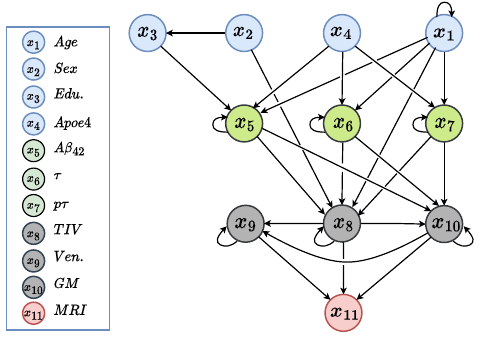}
  \caption{The recovered causal graph, represented by a summary causal graph. Refer to Table~\ref{Shared Edges} for a precise reference to each edge.}
  \label{causal graph}
\end{figure}
\begin{figure}[t]
  \centering
  \includegraphics[width=0.4\textwidth]{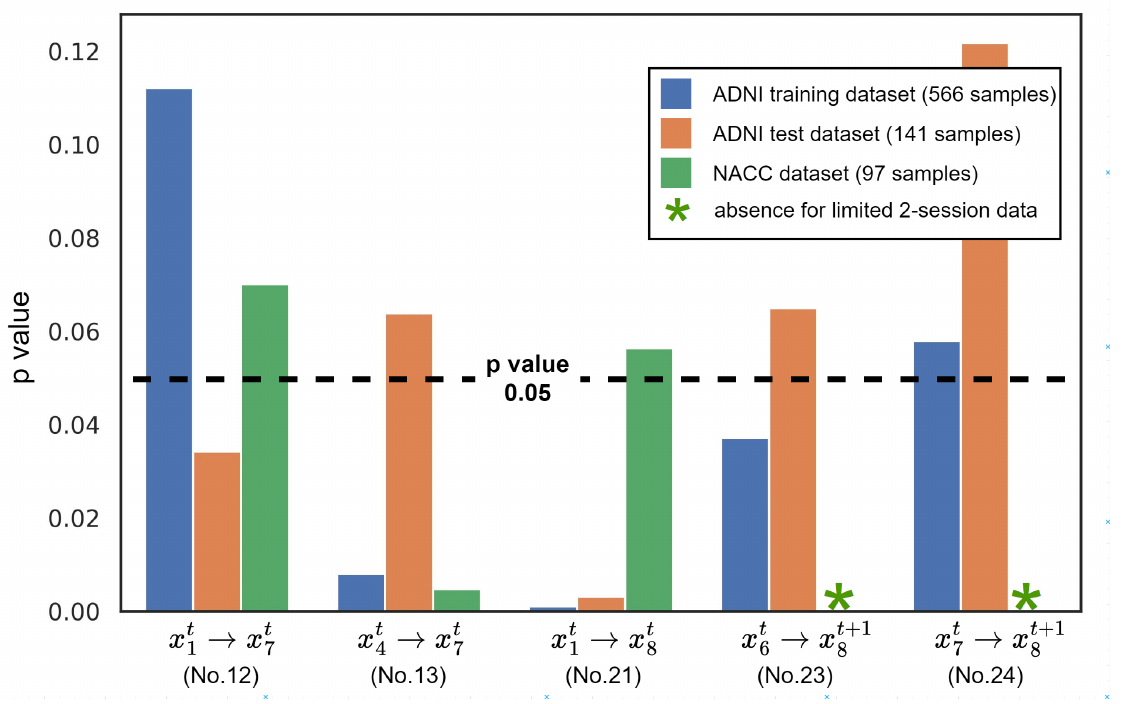}
  \caption{The p-values of edges that do not meet the criteria of t-test in at least one dataset. Each sample of ADNI includes two adjacent session data as $(\boldsymbol{X}^{T}, \boldsymbol{X}^{T+1})$. The NACC dataset lacks multiple session data for a single subject thus only edges of instantaneous causality can be tested. }
  \label{p_value}
\end{figure}
\begin{table}[t]
  \centering
  \caption{The recovered edges.}\label{Shared Edges}\renewcommand{\arraystretch}{1.2}
    \begin{tabular}{llll}
    \hline\hline
    \multicolumn{4}{c}{\textbf{Self-Causality assumption}} \\
    \ 1.$x_5^t\rightarrow x_5^{t+1}$&  \ 2.$x_6^t\rightarrow x_6^{t+1}$&
    \ 3.$x_7^t\rightarrow x_7^{t+1}$&
    \ 4.$x_8^t\rightarrow x_8^{t+1}$\\
    \ 5.$x_9^t\rightarrow x_9^{t+1}$&
    \ 6.$x_{10}^t\rightarrow x_{10}^{t+1}$&       &  \\\midrule
    \multicolumn{4}{c}{\textbf{Recovered by all methods}} \\
    \ 7.$x_2^t\rightarrow x_3^{t}$&
    \ 8.$x_1^t\rightarrow x_5^{t}$&
    \ 9.$x_4^t\rightarrow x_5^{t}$&
     10.$x_1^t\rightarrow x_6^{t}$\\
     11.$x_4^t\rightarrow x_6^{t}$&
     12.$x_1^t\rightarrow x_7^{t}$&
     13.$x_4^t\rightarrow x_7^{t}$&
     14.$x_2^t\rightarrow x_8^{t}$\\
     15.$x_{1}^t\rightarrow x_9^{t}$&
     16.$x_{10}^t\rightarrow x_9^{t}$&
     17.$x_8^t\rightarrow x_{10}^{t}$&
     18.$x_5^t\rightarrow x_{10}^{t+1}$\\
     19.$x_6^t\rightarrow x_{10}^{t+1}$&
     20.$x_7^t\rightarrow x_{10}^{t+1}$\\
    \hline
    \multicolumn{4}{c}{\textbf{Recovered by two methods}}\\
    \multicolumn{2}{l}{21.$x_1^t\rightarrow x_{8}^{t}$ \hspace{1.4em} (FCI, GES)} & \multicolumn{2}{l}{22.$x_5^t\rightarrow x_{8}^{t+1}$ \ (FCI, GES)}\\
    \multicolumn{2}{l}{23.$x_6^t\rightarrow x_{8}^{t+1}$}(GES, LiNGAM)&
    \multicolumn{2}{l}{24.$x_7^t\rightarrow x_{8}^{t+1}$}(GES, LiNGAM)\\
    \multicolumn{2}{l}{25.$x_8^t\rightarrow x_{9}^{t}$ \hspace{1.4em} (FCI, GES)}\\
    \hline\hline
    \end{tabular}
\end{table}%

Fig.~\ref{causal graph} shows the recovered summary causal graph and Table~\ref{Shared Edges} shows the recovered edges in detail. Overall, the introduction of prior knowledge~\ref{PK} ensures that the causal relationships identified do not contradict common sense and existing research findings. For example, there exists no causal edge from $t_1$ to $t_0$, which is contradicted by the time-priority assumptions, or causal edge from the brain tissue volumes to the bio-marker level, which is contradicted with the existing research, as the alternations of proteins is an early event in the progression of AD~\cite{event_order1, event_order2, event_order3}. 

In the partial correlation analysis, 20 of the 25 identified edges demonstrate significance ($\alpha > 0.05$) across the ADNI training set, ADNI validation set, and the NACC dataset. Figure ~\ref{p_value} highlights the edges failing to meet this threshold, along with their respective t-test p-values. Five edges do not meet the criteria, with four (No.12, No.13, No.23, No.24) involving biomarker variables. Interestingly, these edges align with current clinical research. For instance, No.12 and No.13 suggest causal effects of age and the APOE gene on $p\tau$ levels, both recognized as AD risk factors. No.21 indicates age's impact on total intracranial volume (TIV), supported by references ~\cite{age_and_VV1, age_and_VV2, age_and_VV3}. Also, No.23 and No.24 link $\tau$ and $p\tau$ levels to TIV, corroborated by ~\cite{tau_and_VV1, tau_and_VV2}.

The $t$-test's limited success might be attributed to biomarker value fluctuations due to varying measurement techniques. Despite normalization efforts, significant noise affects accuracy.

\underline{SCM fitting validity}:
This part includes assessing the predictive performance of fitted structural causal models. All available data pairs (not necessarily adjacent) are used for structural causal model fitting to capture both short and long-term interval relationships.

Our causality-based model, compared against leading time-series models, shows superior performance in predicting brain volume changes, as depicted in Table~\ref{Attributes result}. For segmented volumes (TIV, VV, and GMV), the data includes 2,762 sessions from 786 subjects (average interval: $2.70_{\pm{2.47}}$ years) and 688 sessions from 186 subjects (average interval: $2.38_{\pm{2.17}}$ years). However, due to limited CSF biomarker data, the training set for CSF biomarkers ($A\beta$, $\tau$, and $p\tau$) consists of 1,231 sessions from 382 subjects (average interval: $3.38_{\pm{2.51}}$ years), and the test set includes 318 sessions from 109 subjects (average interval: $3.11_{\pm{2.13}}$ years).

We utilize Normalized Mean Absolute Error (NMAE) as our metric, defined as $NMAE = \frac{1}{N} \sum_{i=1}^N{\frac{|y_i-\tilde{y_i}|}{y_{max}-y_{min}}}$, where $y_{max}$ and $y_{min}$ are the maximum and minimum values of $y$, $y_i$ is the ground truth, $\tilde{y_i}$ the model prediction, and $N$ the total number of test data.

We compare non-causality-based methods like non-causal linear models, MLPs, RNNs, and LSTMs, which use all variables to predict future outcomes, against causality-based approaches such as causal linear models, causal NFs, and causal MLPs, which only consider causal parent variables for predictions. In all cases except the $A\beta$ variable, causality-based methods outperform their non-causal counterparts. The underperformance with $A\beta$ may stem from its limited causal parent variables providing insufficient information. Notably, even the causal linear method rivals RNNs, LSTMs, and FC Networks in predicting $p\tau$, TIV, VV, and GMV, underscoring how causality enhances model robustness and mitigates spurious correlations.

\begin{figure*}[t]
  \centering
  \includegraphics[width=0.9\textwidth]{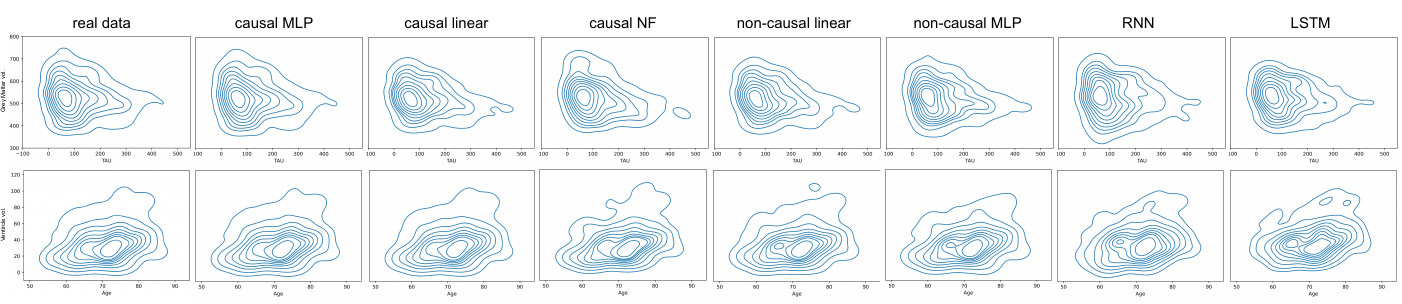}
  \caption{The visualization of the data distribution predicted by different methods. The first row is the joint distribution of $\tau$ (pg/ml) and grey matter volume (ml), while the second row is the joint distribution of age (years) and ventricle volume (ml). The first column is the distribution of real data and the second to eight columns demonstrate the data distribution predicted by different methods. It appears that causal MLP performs the best.}
  \label{fig:data distribution}
\end{figure*}

Figure~\ref{fig:data distribution} illustrates the data distributions from various methods. The first row shows the joint distribution of $\tau$ and GMV, highlighting the negative correlation indicative of $\tau$'s toxicity and its causal effect on GMV atrophy, confirmed by research~\cite{tau_and_GM0, tau_and_GM1, tau_and_GM2}. The second row's positive correlation between age and VV is aligned with findings in~\cite{age_and_VV1, age_and_VV2, age_and_VV3}. The causal MLP most accurately reflects the actual data distribution in both scenarios.

In summary, the causal MLP excels due to the neural network's superior modeling capabilities. However, the causal NF yields higher errors than the causal linear method in low-dimensional settings, suggesting its excessive complexity. This experiment validates our approach of integrating NNs with nonlinear complexity into SCM, effectively addressing {\bf ineffective causal modeling}.

\begin{table}[t]\caption{The predicted non-image variables results of ADNI dataset.}\label{Attributes result}\renewcommand{\arraystretch}{1.2}
  \centering\scalebox{0.85}{
        \begin{tabular}{l|rrrrrr}
    \hline\hline
    \multirow{2}{*}{Methods} & \multicolumn{6}{c}{NMAE$\downarrow$}\\
    \cline{2-7} 
     &$A\beta$ & $\tau$ & $p\tau$ & $TIV$ & $VV$ & $GMV$\\
    \hline
    Non-causal Linear
    &0.0664&0.0662&0.0552&0.0286&0.0273&0.0448\\
    Non-causal MLP
    &0.0653&0.0578&0.0556&0.0197&0.0171&0.0366\\
    RNN
    &{\bf 0.0480}&0.0577&0.0464&0.0333&0.0358&0.0568\\
    LSTM
    &0.0488&0.0565&0.0506&0.0287&0.0276&0.0548\\
    Causal Linear
    &0.0754&0.0720&0.0473&{\bf 0.0181}&0.0173&0.0378\\
    Causal NF
    &0.0917&0.1008&0.0754&0.0292&0.0248&0.0581\\
    Causal MLP
    &0.0518&{\bf 0.0553}&{\bf 0.0404}&0.0182&{\bf 0.0130}&{\bf 0.0326}\\
    \hline\hline
   \end{tabular}}
\end{table}%

\subsubsection{Causal Image Synthesis}
For image synthesis, we utilize StyleGAN, known for its exceptional quality in generating natural images~\cite{karras2019style} and its popularity in image manipulation. We compare StyleGAN with AE-GAN (a variational autoencoder GAN)\cite{vae_gan}, wGAN (a classical Wasserstein GAN with gradient penalty)\cite{wGAN}, and HA-GAN (a hierarchically-amortized GAN for 3D medical image synthesis)~\cite{HA_GAN}.

\underline{Quality of sampled MRI synthesis}: To evaluate the quality of synthesized MRIs, we calculate the Fréchet Inception Distance (FID) and Maximum Mean Discrepancy (MMD), as shown in Table~\ref{FID and MMD}. Both metrics assess the divergence between image distributions, with lower values indicating more realistic MRI simulations. For FID, we use the middle slice from the Sagittal, Coronal, and Axial axes as 2D images and calculate their mean. MMD is computed from a stack of 8 slices.

Except for Latent Diffusion Models (LDM), all methods are trained on the ADNI dataset and tested for FID and MMD against real images from ADNI, OASIS, and NACC datasets. Due to computational constraints, we use a pre-trained LDM checkpoint by MONAI~\cite{monai_gen, brain_LDM}, trained on the UK Biobank (UKB) dataset~\cite{uk_biobank}. For fairness, we remove skulls from LDM output using ROBEX~\cite{robex}.

As Table~\ref{FID and MMD} indicates, StyleGAN achieves the lowest FID and MMD values with fewer parameters and faster inference time. The higher FID and MMD for LDM may be due to dataset differences between UKB and ADNI. Additionally, LDM's slow inference and large parameter count hinder its use in our MRI synthesis module.

\begin{figure}[t]
  \centering
  \includegraphics[width=0.45\textwidth]{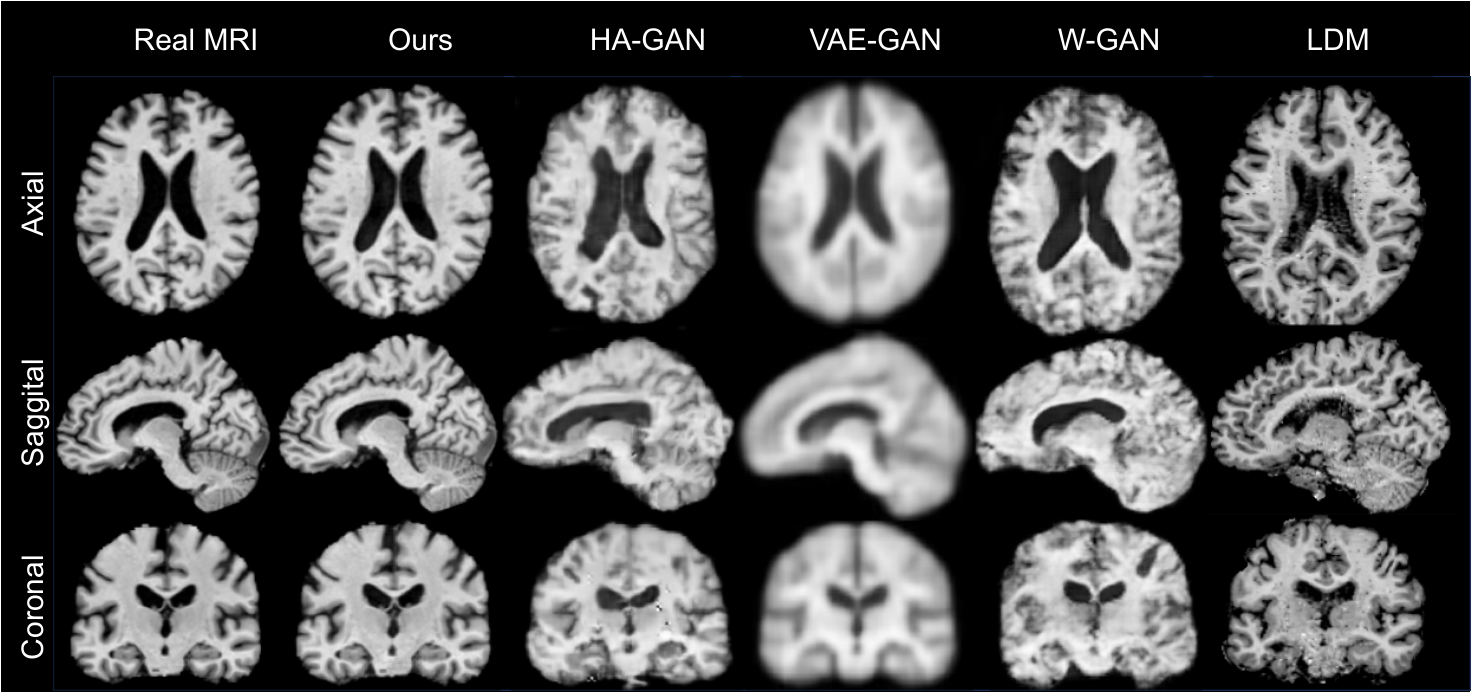}
  \caption{The visualization of the MRI reconstructed by different methods. The first row, second row, and third row respectively show the axial, sagittal, and coronal slice. The first column demonstrates the real MRI and the second to five columns illustrate the MRI reconstructed by different methods.  }\label{fig:rec visualisation}
\end{figure}

MRI reconstruction involves training an encoder to map MRIs to latent space. For W-GAN, HA-GAN, and LDM, we use a ResNet50-based encoder. LDM employs encoded MRI features to guide its denoising process, as described in \cite{}. VAE-GAN utilizes its integrated trained encoder. Our style-based generator’s encoder is detailed in Section\ref{sec:styleGAN's encoder}. As Fig.~\ref{fig:rec visualisation} demonstrates, only StyleGAN produces high-quality MRI reconstructions. VAE-GAN results are blurry, while w-GAN and HA-GAN introduce distortions. LDM reconstructions show substantial noise.

\begin{table}[t]\caption{The performance comparison of sampled MRI quality.}\label{FID and MMD}\renewcommand{\arraystretch}{1.2}\scriptsize
  \centering\scalebox{0.65}{
        \begin{tabular}{l|c|c|c|r r}
    \hline\hline
    \multicolumn{1}{l|}{\multirow{2}{*}{Method}} & \multicolumn{1}{c|}{ADNI} & \multicolumn{1}{c|}{OASIS} & \multicolumn{1}{c|}{NACC} &   
    \multicolumn{1}{l}{\multirow{1}{*}{\#params}} & 
    \multicolumn{1}{l}{\multirow{1}{*}{inf. time}} \\ 
    \cline{2-4}   
     &  \multicolumn{1}{c}
    {FID$\downarrow$/MMD$\downarrow$}& \multicolumn{1}{c}{FID$\downarrow$/MMD$\downarrow$} & \multicolumn{1}{c|}{FID$\downarrow$/MMD$\downarrow$} & \multicolumn{1}{c}{(Mb)} & \multicolumn{1}{c}{(S)} \\
    \hline
    \multicolumn{1}{l|}{VAE-GAN}
    &181.91$_{\pm 0.57}$/6.60$_{\pm 0.01}$
    &189.71$_{\pm 0.72}$/6.91$_{\pm 0.02}$
    &183.94$_{\pm 0.80}$/6.78$_{\pm 0.02}$
    &1739.68&0.088$_{\pm 0.015}$\\
    \multicolumn{1}{l|}{WGAN}
    &167.90$_{\pm 3.43}$/7.73$_{\pm 0.02}$
    &177.14$_{\pm 1.21}$/7.88$_{\pm 0.02}$
    &182.35$_{\pm 1.44}$/7.66$_{\pm 0.03}$
    &481.49&0.094$_{\pm 0.016}$\\
    \multicolumn{1}{l|}{HA-GAN}
    &62.69$_{\pm 0.99}$/5.62$_{\pm 0.01}$
    &80.85$_{\pm 1.86}$/6.36$_{\pm 0.03}$
    &69.27$_{\pm 0.93}$/5.89$_{\pm 0.01}$
    &304.20&0.177$_{\pm 0.054}$\\
    \multicolumn{1}{l|}{LDM}
    &64.87$_{\pm 0.64}$/6.13$_{\pm 0.08}$
    &74.26$_{\pm 1.67}$/6.55$_{\pm 0.09}$
    &56.08$_{\pm 0.34}$/6.24$_{\pm 0.03}$
    &2162.68&7.612$_{\pm 0.228}$\\
    \multicolumn{1}{l|}{Ours}
    &32.90$_{\pm 0.36}$/3.27$_{\pm 0.02}$
    &53.72$_{\pm 0.44}$/3.90$_{\pm 0.06}$
    &48.07$_{\pm 0.41}$/3.33$_{\pm 0.03}$
    &30.06&0.005$_{\pm 0.007}$\\ 
    \multicolumn{1}{l|}{real Data}
    &7.90$_{\pm 0.08}$/1.42$_{\pm 0.06}$
    &19.53$_{\pm 0.25}$/1.78$_{\pm 0.07}$
    &10.46$_{\pm 0.41}$/1.55$_{\pm 0.02}$ & - & - \\ \hline
    \end{tabular}}
\end{table}%

\underline{Quality of volume-intervened MRI synthesis}: 
We test the performance of ISM in generating brain MRIs that intervened on specific brain volumes, including VV, GMV, and TIV. Desired volume changes are set between $+15\%$ and $-15\%$, and the intervened images are analyzed using ANTs~\cite{ANTs} to confirm actual volume alterations. Table~\ref{control volume results} presents the average and standard deviation of the actual volume changes, derived from the segmented volumes of the synthesized MRIs.

\begin{figure*}[t]
  \centering
  \includegraphics[width=0.9\textwidth]{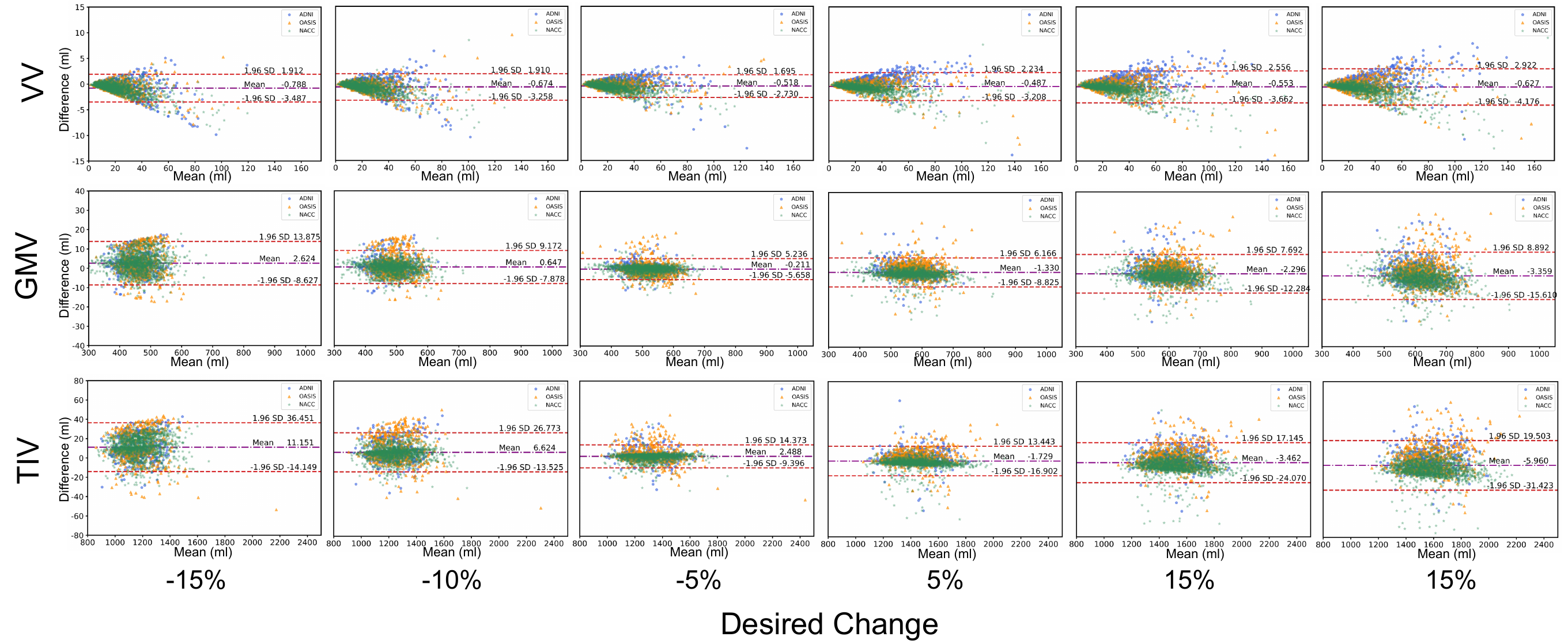}
  \caption{The Bland-Altman plot to illustrate the disparity between the desired volumes and the synthesized MR volumes. The first row illustrates the VV, the second row illustrates the GMV and the third row illustrates the TIV. From left to right, the column represents the desired change percentage from -15\% to 15\%. In each sub-figure, the x-axis represents the mean value of the desired volume and the synthesized MR volume while the y-axis represents the difference between the desired volume and the synthesized MR volume.}
  \label{fig:BAplot}
\end{figure*}

Fig.~\ref{fig:BAplot} features a Bland-Altman plot demonstrating the discrepancy between desired and actual volumes in the synthesized MRIs. The plot's x-axis indicates the average of the desired and synthesized MRI volumes, while the y-axis shows their difference. Different datasets are denoted by varying dot types.

From Table~\ref{control volume results} and Fig.\ref{fig:BAplot}, it is evident that the ADNI-trained model can accurately synthesize MRIs with interventions across various datasets. The maximum error observed was a 3.27\% deviation in MRI with a $15\%$ increased ventricle volume on the OASIS dataset. The stability of outputs tends to decrease with larger desired changes, especially at the extremes of $+15\%$ and $-15\%$. This trend is particularly noticeable for VV interventions, where errors grow as the original VV increases. The first row of Fig.\ref{fig:BAplot} illustrates this, showing a larger and more dispersed difference as the desired volume diverges from the original.

\begin{figure}[t]
  \centering
  \includegraphics[width=0.49\textwidth]{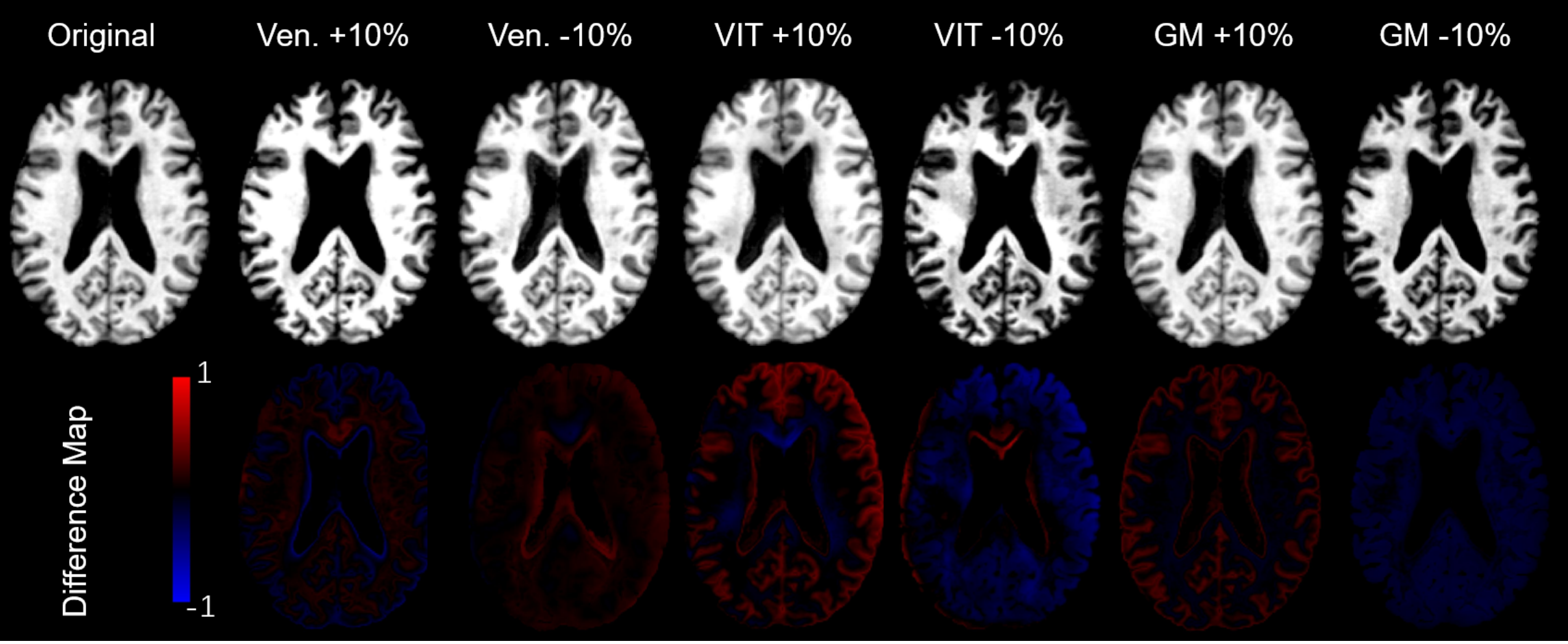}\label{visualise2}
  \caption{The visualization of the synthesized MRI intervened on different volumes. The first row shows the axial slice of the MRI and the second row includes the corresponding difference map w.r.t. the original MRI in the first column.}
  \label{fig:intervene visualise}
\end{figure}

\begin{table}[t]\caption{The performance of volume intervention.}\label{control volume results}\renewcommand{\arraystretch}{1.2}\scriptsize
  \centering
        \begin{tabular}{c|r|r r r}
    \hline\hline
    \multicolumn{1}{c}{\multirow{2}{*}{Dataset}} & \multicolumn{1}{c|}{Desired} & \multicolumn{3}{c}{Actual Change} \\
    \cline{3-5}    \multicolumn{1}{l}{} &    Change (\%)   & VV (\%)&GMV(\%)& TIV(\%)\\
    \hline
    \multirow{5}[2]{*}{ADNI} & -15 &
    -17.54$_{\pm 3.51}$&-15.01$_{\pm 1.38}$&-13.44$_{\pm 2.66}$\\
     & -10 & -11.39$_{\pm 3.19}$ & -9.32$_{\pm 1.21}$& -9.24$_{\pm 1.05}$\\
     & -5 & -5.95$_{\pm 1.29}$ & -4.31$_{\pm0.52}$& -5.36$_{\pm 1.06}$\\
     & 5 & 6.28$_{\pm 1.49}$& 6.31$_{\pm1.05}$ & 6.08$_{\pm 1.08}$ \\
     & 10 & 11.38$_{\pm 2.41}$ & 9.01$_{\pm 1.61}$ & 8.12$_{\pm 2.41}$\\
     & 15 &17.51$_{\pm 3.45}$ & 13.66$_{\pm 1.74}$ & 12.81$_{\pm 2.62}$\\
     \cline{1-5}  
        \multirow{5}[2]{*}{OASIS} & -15 &
    -13.88$_{\pm 2.06}$ & -14.02$_{\pm 2.44}$& -13.97$_{\pm 2.13}$\\
     & -10 & -12.71$_{\pm 1.24}$ & -12.98$_{\pm 1.69}$ & -11.80$_{\pm 1.40}$\\
     & -5 & -5.73$_{\pm 1.33}$ & -5.19$_{\pm 1.67}$ & -5.11$_{\pm 1.56}$ \\
     & 5 & 4.43$_{\pm 1.43}$ & 6.31$_{\pm 1.63}$ & 5.43$_{\pm 1.34}$ \\
     & 10 & 10.61$_{\pm 1.73}$ & 11.99$_{\pm 1.39}$ & 11.10$_{\pm 1.85}$\\
     & 15 & 18.27$_{\pm 2.46}$ & 14.26$_{\pm 1.78}$ & 14.75$_{\pm 1.99}$\\
      \cline{1-5}  
    \multirow{5}[2]{*}{NACC} & -15 &
    -16.94$_{\pm 2.37}$ & -13.94$_{\pm 2.24}$ & -14.00$_{\pm 2.49}$\\
     &-10\ &
     -12.34$_{\pm 2.34}$ & -8.06$_{\pm 2.02}$ & -8.38$_{\pm 1.64}$\\
     & -5 &
    -3.35$_{\pm 1.99}$ & -5.18$_{\pm 1.73}$& -3.77$_{\pm 1.47}$\\
     & 5 &
      5.57$_{\pm 1.30}$ & 4.02$_{\pm 1.20}$& 3.75$_{\pm 2.08}$\\
     & 10 &
     11.20$_{\pm 2.58}$ & 9.21$_{\pm 2.10}$& 9.22$_{\pm 2.22}$\\
     & 15 &
     13.58$_{\pm 3.43}$& 14.36$_{\pm 2.19}$& 13.19$_{\pm 2.66}$ \\
    \hline\hline
    \end{tabular}
\end{table}%

Fig.~\ref{fig:intervene visualise} displays visualizations of volume-intervened MRI. The difference map ${I}{diff} = {I}{ivt} - {I}_{ori}$ illustrates volume changes, with blue indicating decreased and red increased brain tissue intensity. For example, the second column shows a 10\% ventricle enlargement, evidenced by blue around the ventricle. Interventions on total intracranial volume are demonstrated in the third and fourth columns, with red indicating tissue addition and blue indicating reduction, corresponding to the desired changes.

\underline{Longitudinal MRI synthesis}: We evaluate our CLIS model's performance in generating future session images with  M${12}$ (the 12-th month from baseline), M${24}$ (the 24-th month from baseline),  and M$_{48}$ (the 48-th month from baseline) using ADNI and NACC datasets. Our method is compared against image translation methods Pix2Pix~\cite{pix2pix}, CycleGAN~\cite{cycleGAN}, and MRI-specific networks MI-GAN~\cite{multi-information-GAN}, LD-GAN~\cite{LDGAN}, and TR-GAN~\cite{TR-GAN}.

MI-GAN~\cite{multi-information-GAN} is a Multi-Information GAN, which estimates brain MRI at future time-point conditioning on the brain MRI at baseline time-point and multiple information: gender, education level, and APOE gene. LD-GAN~\cite{LDGAN} is a Longitudinal-Diagnostic GAN, which imputes MRIs by learning a bi-directional mapping between MRIs of two adjacent time points and performing clinical score prediction jointly, thereby explicitly encouraging task-oriented image synthesis. TR-GAN~\cite{TR-GAN} is a Temporal Recurrent GAN to complete missing sessions of MRI datasets and adopts recurrent connections to deal with variant input sequence lengths and flexibly generate future sessions. 

For generating the future M$_{12}$, M$_{24}$ and M$_{48}$ images, our model only takes the baseline session and predicts all the multiple future sessions in a universal way, while the Pix2Pix, CycleGAN, and MI-GAN models cannot deal with the varying intervals so we train 3 models for them separately. The TR-GAN and LD-GAN models are designed for multiple inputs, however, for comparison fairness, the model also only takes the baseline session as input.

Quantitative results, including mean absolute error (MAE), multi-sacle structural similarity (SSIM) index, and Peak Signal-to-Noise Ratio (PSNR), are detailed in Tables~\ref{image quality result} and ~\ref{image segment result}. The $t$-test is performed for significance analysis with * for p \textless  0.05, ** for p \textless 0.01, and *** for p \textless 0.001. Our method consistently outperforms other models across metrics and datasets, particularly for M${24}$ and M${48}$ predictions, highlighting the significance of incorporating multi-session relationships.

Table~\ref{image segment result} compares volumetric feature performance. Our model demonstrates the closest match to actual MRI volumes, indicating its superior accuracy in predicting VV, GMV, and TIV changes. The $t$-test is performed for significance analysis with * for p \textless  0.05, ** for p \textless 0.01, and *** for p \textless 0.001. All models show increased errors on the NACC test dataset due to population differences, yet our model exhibits better generalization and less performance degradation.

As NACC only serves as a test dataset, and there exists a population shift compared with the ADNI training dataset; hence all models exhibit poorer performance on the NACC dataset, with larger errors in the volumes of generated MRI. However, compared with other methods, our model demonstrates better generalization, with less performance degradation. As Table~\ref{image segment result} shows, the largest performance degradation of our model happens in the M$_{24}$ synthesis task, as the MAE of total intracranial volume increases from 13.99ml of ADNI to 21.53ml of NACC. On the other hand, while TR-GAN outputs the least error among the other methods, in terms of total intracranial volume, the gap between the MAE of TR-GAN and our model is 10.13/8.48/5.18 in ADNI for M$_{12}$/M$_{24}$/M$_{48}$ MRI synthesis, the gap increases to 15.47/13.99/15.56 in NACC dataset, which confirms the better generalization of our model.

\begin{figure*}[t]
  \centering
  \includegraphics[width=0.9\textwidth]{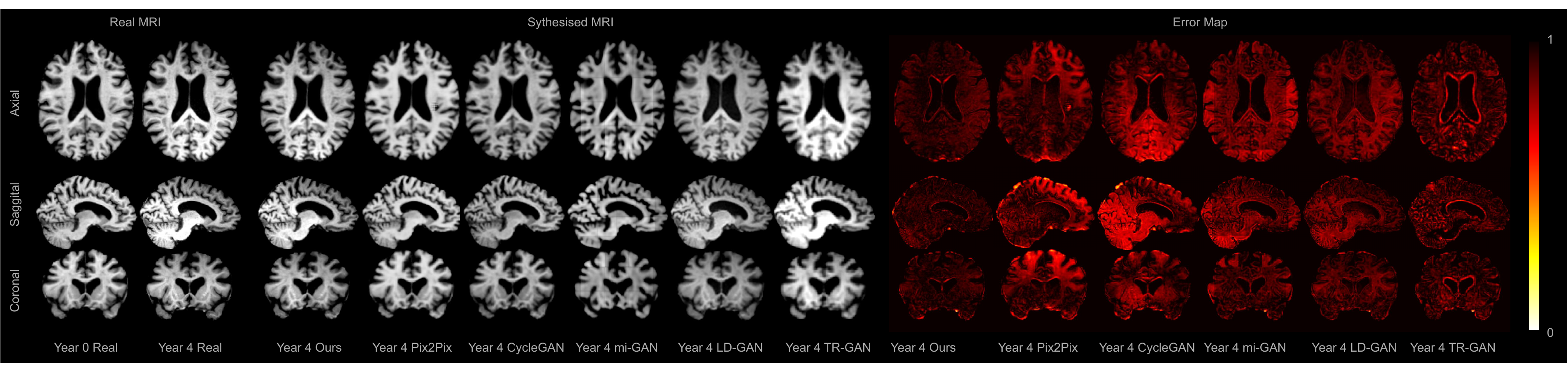}
  \caption{The visualisation of the synthesized MRI by different methods. Based on the MRI at M$_{0}$, the methods synthesize MRI in M$_{48}$ (4 years).  Axial, sagittal, and coronal slices are displayed from the top row to the bottom. The error map represents the absolute error per voxel, on a scale from 0 to 1. In this map, higher values (indicated by brighter colors on the color bar) signify greater error.}
  \label{fig:visualise}
\end{figure*}

Fig.~\ref{fig:visualise} presents the MRI visualization and error maps from different methods. Our method yields high-quality MRIs with few artifacts and blurs, closely matching the target MRI. In contrast, MI-GAN produces MRIs with inconsistent patches, while Pix2Pix and CycleGAN results are generally blurry. TR-GAN and LD-GAN exhibit incorrect tissue contrast, further highlighting the superior accuracy and generalization of our model.

\begin{table*}[t]\caption{The quantitative performance of the longitudinally synthesized images. M$_{n}$ denote the n-th month from baseline. e.g., M$_{00}$$\rightarrow$M$_{12}$ means the synthesis of MRI at 12-th month on the basis of the MRI at 0-th month.}\label{image quality result}\renewcommand{\arraystretch}{1.2}\scriptsize
  \centering\scalebox{0.98}{
        \begin{tabular}{l|l|p{1.4cm}p{1cm}p{1.4cm}|p{1.4cm}p{1cm}p{1.4cm}|p{1.4cm}p{1cm}p{1.4cm}}
    \hline\hline
    \multicolumn{1}{l}{\multirow{2}{*}{Dataset}} & \multicolumn{1}{l|}{\multirow{2}{*}{Method}} & \multicolumn{3}{c|}{M$_{00}$$\rightarrow$M$_{12}$ test:131/217 sub} & \multicolumn{3}{c|}{M$_{00}$$\rightarrow$M$_{24}$
    test:91/417 sub} & \multicolumn{3}{c}{M$_{00}$$\rightarrow$M$_{48}$
    test:41/165 sub} \\
\cline{3-11}    \multicolumn{1}{l}{} &       & MAE$\downarrow$ & SSIM(\%)$\uparrow$ & PSNR(dB)$\uparrow$&MAE$\downarrow$& SSIM(\%)$\uparrow$& PSNR(dB)$\uparrow$&MAE$\downarrow$& SSIM(\%)$\uparrow$&PSNR(dB)$\uparrow$\\
    \hline
    \multirow{5}[2]{*}{ADNI} & Pix2Pix~\cite{pix2pix} &
    0.020$_{\pm0.005}$***&94.9$_{\pm1.7}$***&26.06$_{\pm1.01}$***&
    0.023$_{\pm0.006}$***&94.3$_{\pm1.9}$***&25.61$_{\pm1.79}$***&
    0.027$_{\pm0.008}$***&92.9$_{\pm2.8}$***&23.62$_{\pm1.85}$***\\
        & CycleGAN~\cite{cycleGAN} & 
    0.014$_{\pm0.002}$***&95.9$_{\pm0.7}$***&26.39$_{\pm0.89}$***&
    0.019$_{\pm0.003}$***&93.3$_{\pm1.1}$***&25.76$_{\pm1.02}$***&
    0.024$_{\pm0.004}$***&90.8$_{\pm1.4}$***&24.08$_{\pm1.63}$***\\
    & MI-GAN~\cite{multi-information-GAN}     &
    0.016$_{\pm0.007}$***&95.4$_{\pm0.6}$***&28.33$_{\pm1.23}$***&
    0.043$_{\pm0.006}$***&92.4$_{\pm1.9}$***&19.02$_{\pm1.66}$***&
    0.042$_{\pm0.004}$***&92.2$_{\pm0.9}$***&19.20$_{\pm1.76}$***\\
    & LD-GAN~\cite{LDGAN} &
    0.013$_{\pm0.003}$***&97.5$_{\pm0.7}$***&29.11$_{\pm1.65}$***&
    0.015$_{\pm0.003}$***&96.3$_{\pm1.2}$***&28.27$_{\pm1.74}$***&
    0.021$_{\pm0.004}$***&95.5$_{\pm1.2}$***&26.68$_{\pm2.10}$***\\
    &  TR-GAN~\cite{TR-GAN}     &                       
    0.012$_{\pm0.010}$***&97.1$_{\pm4.3}$**&29.12$_{\pm1.90}$***&
    0.012$_{\pm0.008}$*&97.3$_{\pm2.9}$*&29.05$_{\pm1.48}$**&
    0.014$_{\pm0.011}$*&96.8$_{\pm3.4}$*&27.82$_{\pm2.32}$***\\
&   CLIS (ours)    &
    \textbf{0.008$_{\pm0.002}$}&\textbf{98.4$_{\pm0.9}$}&\textbf{31.33$_{\pm1.81}$}&
    \textbf{0.009$_{\pm0.003}$}&\textbf{98.2$_{\pm1.4}$}&\textbf{30.71$_{\pm2.22}$}&
    \textbf{0.011$_{\pm0.003}$}&\textbf{97.8$_{\pm 1.5}$}&\textbf{29.76$_{\pm2.23}$}\\
    \hline
    \multirow{5}[2]{*}{NACC} & Pix2Pix~\cite{pix2pix} & 0.021$_{\pm0.008}$***&94.3$_{\pm3.2}$***&25.64$_{\pm1.83}$***&
    0.025$_{\pm0.008}$***&94.1$_{\pm2.2}$***&24.85$_{\pm2.67}$***&
    0.026$_{\pm0.015}$***&92.7$_{\pm2.9}$***&22.84$_{\pm2.81}$***\\
    & CycleGAN~\cite{cycleGAN} &
    0.016$_{\pm0.004}$***&95.3$_{\pm1.8}$***&25.67$_{\pm1.52}$***&
    0.022$_{\pm0.004}$***&94.7$_{\pm2.3}$***&25.29$_{\pm2.08}$***&
    0.027$_{\pm0.005}$***&91.8$_{\pm2.8}$***&21.73$_{\pm2.09}$***\\
    & MI-GAN~\cite{multi-information-GAN}&
    0.031$_{\pm0.004}$***&92.9$_{\pm1.3}$***&21.19$_{\pm1.53}$***&
    0.044$_{\pm0.008}$***&92.7$_{\pm3.1}$***&18.39$_{\pm1.67}$***&
    0.043$_{\pm0.010}$***&92.2$_{\pm4.3}$***&18.76$_{\pm2.45}$***\\
    & LD-GAN~\cite{LDGAN} & 
    0.020$_{\pm0.005}$***&95.8$_{\pm1.9}$***&26.78$_{\pm2.74}$***&
    0.022$_{\pm0.003}$***&95.1$_{\pm2.4}$***&26.23$_{\pm2.72}$***&
    0.024$_{\pm0.006}$***&94.0$_{\pm3.1}$***&25.52$_{\pm3.03}$*** \\
    &   TR-GAN~\cite{TR-GAN}    &             
    0.014$_{\pm0.010}$***&96.2$_{\pm4.8}$*&27.82$_{\pm4.11}$***&
    0.013$_{\pm0.005}$*&96.5$_{\pm2.9}$*&27.84$_{\pm2.73}$***&
    0.015$_{\pm0.006}$**&95.6$_{\pm3.4}$**&26.66$_{\pm2.80}$***\\
    & CLIS (ours)    &
    \textbf{0.011$_{\pm0.005}$}&\textbf{97.3$_{\pm2.7}$}&\textbf{29.37$_{\pm3.02}$}&
    \textbf{0.012$_{\pm0.005}$}&\textbf{97.0$_{\pm2.6}$}&\textbf{28.86$_{\pm2.94}$}&
    \textbf{0.012$_{\pm0.006}$}&\textbf{96.8$_{\pm2.7}$}&\textbf{28.55$_{\pm2.93}$}\\
    \hline\hline
    \end{tabular}}
\end{table*}%

\begin{table*}[t]\caption{The performance comparison in  terms of volumetric features.}\label{image segment result}\renewcommand{\arraystretch}{1.2}\scriptsize
  \centering\scalebox{0.93}{
        \begin{tabular}{l|l|c c c|c c c|c c c}
    \hline\hline
    \multicolumn{1}{l}{\multirow{2}{*}{Methods}} & \multicolumn{1}{l|}{\multirow{2}{*}{Dataset}} & \multicolumn{3}{c|}{M$_{00}$$\rightarrow$M$_{12}$ test:131/217 sub} & \multicolumn{3}{c|}{M$_{00}$$\rightarrow$M$_{24}$
    test:91/417 sub} & \multicolumn{3}{c}{M$_{00}$$\rightarrow$M$_{48}$
    test:41/165 sub} \\
\cline{3-11}    \multicolumn{1}{l}{} &       & \multicolumn{9}{c}{MAE of VV/ GMV / TIV (ml) $\downarrow$}\\
    \hline
    \multirow{5}[2]{*}{ADNI} & Pix2Pix & 
    \multicolumn{3}{c|}
    {5.63$_{\pm1.81}$*** / 25.36$_{\pm4.58}$*** / 23.76$_{\pm4.95}$***}&
    \multicolumn{3}{c|}
    {5.94$_{\pm1.80}$*** / 29.32$_{\pm4.40}$*** / 25.25$_{\pm6.85}$***}&
    \multicolumn{3}{c}
    {7.28$_{\pm1.85}$*** / 46.44$_{\pm4.66}$*** / 26.17$_{\pm4.68}$***}\\
     &CycleGAN & \multicolumn{3}{c|}
    {8.35$_{\pm1.20}$*** / 24.05$_{\pm4.50}$*** / 31.09$_{\pm4.41}$***}&
    \multicolumn{3}{c|}
    {9.92$_{\pm2.08}$*** / 24.03$_{\pm4.67}$*** / 28.52$_{\pm4.74}$***}&
    \multicolumn{3}{c}
    {\!10.22$_{\pm2.74}$***/ 37.69$_{\pm6.21}$*** / 30.67$_{\pm5.45}$***}\\
     & MI-GAN &\multicolumn{3}{c|}
     {2.74$_{\pm1.36}$**\quad / 22.75$_{\pm3.65}$*** / 50.51$_{\pm5.25}$***}&
    \multicolumn{3}{c|}
    {4.76$_{\pm1.45}$*** / 86.20$_{\pm5.02}$*** / 67.20$_{\pm5.49}$***}&
    \multicolumn{3}{c}
    {6.51$_{\pm2.43}$*** / 76.95$_{\pm6.09}$*** / 67.92$_{\pm8.91}$***}\\
    & LD-GAN &
    \multicolumn{3}{c|}
    {3.38$_{\pm1.96}$*** / 28.41$_{\pm3.95}$*** / 30.68$_{\pm7.27}$***}&
    \multicolumn{3}{c|}
    {3.33$_{\pm1.95}$*\quad \ / 43.75$_{\pm6.64}$*** / 43.02$_{\pm7.80}$***}&
    \multicolumn{3}{c}
    {6.37$_{\pm2.53}$*** / 46.93$_{\pm6.71}$*** / 32.16$_{\pm6.90}$***}
\\
     &TR-GAN  & 
    \multicolumn{3}{c|}
    {3.01$_{\pm1.45}$*** / 14.48$_{\pm3.66}$*\quad \ / 23.90$_{\pm3.43}$***}&
    \multicolumn{3}{c|}
    {3.26$_{\pm2.78}$*\quad \ / 16.94$_{\pm2.87}$*\quad \ / 22.47$_{\pm4.99}$***}&
    \multicolumn{3}{c}
    {4.38$_{\pm2.08}$*\quad \ / 23.34$_{\pm4.73}$*\quad \ / 23.21$_{\pm4.16}$***}\\
      & CLIS (ours) & 
    \multicolumn{3}{c|}
    {\textbf{2.27$_{\pm1.18}$} \quad \; / \textbf{13.57$_{\pm2.34}$} \quad \;/ \textbf{13.77$_{\pm2.87}$} \quad \: }&
    \multicolumn{3}{c|}
    {\textbf{2.87$_{\pm1.14}$} \quad \; / \textbf{16.28$_{\pm2.81}$} \quad \;/ \textbf{13.99$_{\pm2.54}$} \quad \;}&
    \multicolumn{3}{c}
    {\textbf{3.63$_{\pm1.81}$} \quad \; / \textbf{22.88$_{\pm2.27}$} \quad \;/ \textbf{18.03$_{\pm3.27}$} \quad \;}\\
    \hline
    \multirow{5}[2]{*}{NACC} & Pix2Pix &
     \multicolumn{3}{c|}
     {5.63$_{\pm1.67}$*** / 53.05$_{\pm5.15}$*** / 28.05$_{\pm2.31}$***}&
     \multicolumn{3}{c|}
     {6.64$_{\pm1.98}$*** / 29.73$_{\pm4.00}$*** / 31.46$_{\pm5.53}$***}&
     \multicolumn{3}{c}
     {6.93$_{\pm2.23}$*** / 37.74$_{\pm4.45}$*** / 40.29$_{\pm6.58}$***}\\
    &CycleGAN&    
    \multicolumn{3}{c|}
    {8.79$_{\pm1.88}$*** / 25.66$_{\pm5.51}$*** / 32.03$_{\pm6.99}$***}&
    \multicolumn{3}{c|}
    {7.43$_{\pm1.97}$*** / 24.72$_{\pm4.49}$*** / 35.08$_{\pm8.29}$***}&
    \multicolumn{3}{c}
    {7.35$_{\pm2.31}$*** / 30.69$_{\pm4.83}$*** / 34.61$_{\pm9.28}$***}\\
     & MI-GAN&
    \multicolumn{3}{c|}{3.69$_{\pm5.33}$*** / 25.55$_{\pm2.31}$*** / 46.97$_{\pm2.16}$***}&
    \multicolumn{3}{c|}{3.12$_{\pm2.81}$*** / 23.32$_{\pm2.24}$*** / 42.27$_{\pm4.89}$***}&
    \multicolumn{3}{c} {3.69$_{\pm5.33}$*** / 25.55$_{\pm2.31}$*** / 46.97$_{\pm2.16}$***}\\
    & LD-GAN &
    \multicolumn{3}{c|}{3.19$_{\pm1.76}$*\quad \ / 27.44$_{\pm3.65}$*** / 30.74$_{\pm7.15}$***}&
    \multicolumn{3}{c|}{3.33$_{\pm1.95}$*\quad \ / 43.75$_{\pm6.64}$*** / 43.02$_{\pm7.80}$***}&
    \multicolumn{3}{c}{4.79$_{\pm2.70}$*** / 44.88$_{\pm6.35}$*** / 41.04$_{\pm8.68}$***}\\
    &  TR-GAN     &             
    \multicolumn{3}{c|}{4.05$_{\pm1.69}$*\quad \ / 23.31$_{\pm1.79}$*\quad \  / 35.73$_{\pm3.01}$***}&
    \multicolumn{3}{c|}{3.30$_{\pm2.54}$*\quad \ / 22.74$_{\pm2.12}$*\quad \ / 35.52$_{\pm2.75}$***}&
    \multicolumn{3}{c} {3.41$_{\pm3.19}$  \quad \;/ 30.30$_{\pm3.16}$*** / 38.75$_{\pm4.24}$***}\\
    &  CLIS (ours)     &
    \multicolumn{3}{c|}{\textbf{2.43$_{\pm1.96}$} \quad \; / \textbf{22.95$_{\pm2.37}$ \quad \;/ 20.26$_{\pm2.46}$} \quad \;}&
    \multicolumn{3}{c|}{\textbf{2.83$_{\pm1.54}$} \quad \; / \textbf{21.99$_{\pm1.36}$} \quad \;/ \textbf{21.53$_{\pm3.35}$} \quad \;}&
    \multicolumn{3}{c}{\textbf{3.28$_{\pm1.70}$} \quad \; / \textbf{26.88$_{\pm1.49}$} \quad \;/ \textbf{23.19$_{\pm2.88}$} \quad \;}\\
    \hline\hline
    \end{tabular}}
\end{table*}%

\subsubsection{Downstream Tasks of AD Characterization}\label{sec:downstream}
To establish the clinical relevance of our generated images, we apply them to predict future disease progression in two tasks: AD classification and ventricle volume prediction.

Firstly, we conduct a three-class classification task (AD vs. NC vs. MCI) using ADNI and NACC datasets, with results presented in Table~\ref{classification}. We assess performance using multi-class AUC (mAUC), accuracy, F1 score, recall, and precision. The synthesized MRIs from our model prove comparable to real MRIs, surpassing other methods in almost all metrics across scenarios. Notably, our model demonstrates superior performance in the M$_{48}$ task, the largest time interval, which is crucial yet challenging in clinical settings. The NACC dataset, having a higher proportion of NC cases, show that our model's advantage is less pronounced compared to ADNI.

Secondly, our method is applied in the TADPOLE Challenge for VV prediction, competing with 58 other algorithms. Our test set includes 696 sessions from 335 subjects, with an average interval of $3.01_{\pm 1.01}$ years. Our model achieves an MAE of 0.39, outperforming the winning algorithm's 0.41, without utilizing additional PET and DTI data available in the TADPOLE. We anticipate that incorporating PET and DTI data could further enhance our results.

\begin{table*}[t]\caption{The performance comparison in terms of AD classification.}\label{classification}\renewcommand{\arraystretch}{1.2}\scriptsize
  \centering
        \begin{tabular}{l|l|c c c|c c c|c c c}
    \hline\hline
    \multicolumn{1}{l}{\multirow{2}{*}{Methods}} & \multicolumn{1}{l|}{\multirow{2}{*}{Dataset}} & \multicolumn{3}{c|}{M$_{00}$$\rightarrow$M$_{12}$ test:131/217 sub} & \multicolumn{3}{c|}{M$_{00}$$\rightarrow$M$_{24}$
    test:91/417 sub} & \multicolumn{3}{c}{M$_{00}$$\rightarrow$M$_{48}$
    test:41/165 sub} \\
\cline{3-11}    \multicolumn{1}{l}{} &       & \multicolumn{9}{c}{mAUC / acc / F1 / recall / precision}\\
    \hline
    \multirow{5}[2]{*}{ADNI} & Pix2Pix & 
    \multicolumn{3}{c|}
    {0.916 / 0.782 / 0.594 / 0.648 / 0.553}&
    \multicolumn{3}{c|}
    {0.833 / 0.769 / 0.566 / 0.617 / 0.526}&
    \multicolumn{3}{c}
    {0.769 / 0.700 / 0.508 / 0.558 / 0.473}\\
    & CycleGAN & 
    \multicolumn{3}{c|}
    {0.897 / 0.655 / 0.663 / 0.680 / 0.750}&
    \multicolumn{3}{c|}
    {0.831 / 0.629 / 0.578 / 0.581 / 0.696}&
    \multicolumn{3}{c}
    {0.754 / 0.583 / 0.452 / 0.467 / 0.700}\\
    & MI-GAN  &
    \multicolumn{3}{c|}{0.797 / 0.660 / 0.492 / 0.576 / 0.438}&
    \multicolumn{3}{c|}{0.613 / 0.538 / 0.392 / 0.469 / 0.389}&
    \multicolumn{3}{c}{0.527 / 0.411 / 0.276 / 0.379 / 0.317}\\
    &LD-GAN&
    \multicolumn{3}{c|}{0.820 / 0.811 / 0.732 / 0.718 / 0.814}&
    \multicolumn{3}{c|}{0.712 / 0.621 / 0.600 / 0.621 / 0.715}&
    \multicolumn{3}{c}{0.676 / 0.622 / 0.561 / 0.567 / 0.683}\\
    & TR-GAN & 
    \multicolumn{3}{c|}
    {0.913 / 0.785 / 0.608 / 0.654 / 0.888}&
    \multicolumn{3}{c|}
    {0.848 / 0.769 / 0.566 / 0.617 / 0.526}&
    \multicolumn{3}{c}
    {0.805 / 0.700 / 0.508 / 0.558 / 0.473}\\
    & CLIS (ours)  & 
    \multicolumn{3}{c|}
    {\textbf{0.926} / \textbf{0.881} / \textbf{0.848} / \textbf{0.826} / \textbf{0.906}}&
    \multicolumn{3}{c|}
    {\textbf{0.865} / \textbf{0.804} / \textbf{0.727} / \textbf{0.712} / \textbf{0.810}}&
    \multicolumn{3}{c}
    {\textbf{0.825} / \textbf{0.717} / \textbf{0.573} / \textbf{0.592} / \textbf{0.813}}\\
\cline{3-11}    
    & Real Images & 
    \multicolumn{3}{c|}
    {0.985 / 0.889 / 0.855 / 0.835 / 0.910}&
    \multicolumn{3}{c|}
    {0.956 / 0.832 / 0.747 / 0.735 / 0.826}&
    \multicolumn{3}{c}
    {0.914 / 0.783 / 0.617 / 0.639 / 0.847}\\
    \hline
     \multirow{5}[2]{*}{NACC} & Pix2Pix &
     \multicolumn{3}{c|}
     {\textbf{0.865} / 0.802 / 0.779 / 0.772 / 0.805}&
     \multicolumn{3}{c|}
     {\textbf{0.878} / 0.883 / 0.802 / 0.754 / 0.881}&
     \multicolumn{3}{c}
     {0.836 / 0.828 / 0.720 / 0.690 / 0.786}\\
      & CycleGAN &
     \multicolumn{3}{c|}
     {0.857 / 0.763 / 0.725 / 0.704 / 0.814}&
     \multicolumn{3}{c|}
     {0.865 / 0.847 / 0.744 / 0.685 / 0.857}&
     \multicolumn{3}{c}
     { 0.809 / 0.816 / 0.692 / 0.651 / 0.769}\\
    & MI-GAN&
    \multicolumn{3}{c|}{0.647 / 0.689 / 0.571 / 0.542 / 0.762}&
    \multicolumn{3}{c|}{0.585 / 0.718 / 0.464 / 0.447 / 0.816}&
    \multicolumn{3}{c} {0.530 / 0.693 / 0.346 / 0.371 / 0.721}
    \\
    &   LD-GAN    &             
    \multicolumn{3}{c|}{0.856 / 0.809 / 0.788 / 0.783 / 0.812}&
    \multicolumn{3}{c|}{0.804 / 0.809 / 0.788 / \textbf{0.783} / 0.812}&
    \multicolumn{3}{c}{0.797 / 0.833 / 0.732 / 0.715 / 0.791}
    \\
    &   TR-GAN    &             
    \multicolumn{3}{c|}{0.862 / 0.784 / 0.754 / 0.743 / 0.787}&
    \multicolumn{3}{c|}{0.871 / 0.890 / 0.812 / 0.769 / \textbf{0.885}}&
    \multicolumn{3}{c} {0.825 / 0.828 / 0.724 / 0.701 / 0.786}\\
    &   CLIS (ours)    &
    \multicolumn{3}{c|}{0.863 / \textbf{0.813} / \textbf{0.792} / \textbf{0.785} / \textbf{0.816}}&
    \multicolumn{3}{c|}{0.866 / \textbf{0.890} / \textbf{0.812} / 0.769 / \textbf{0.885}}&
    \multicolumn{3}{c} {\textbf{0.844} / \textbf{0.845} / \textbf{0.744} / \textbf{0.720} / \textbf{0.805}}\\
\cline{3-11}     
    &   Real Images    &
    \multicolumn{3}{c|}{0.881 / 0.831 / 0.808 / 0.799 / 0.837}&
    \multicolumn{3}{c|}{0.900 / 0.911 / 0.833 / 0.785 / 0.913}&
    \multicolumn{3}{c} {0.847 / 0.885 / 0.779 / 0.739 / 0.854}\\ 
    \hline\hline
    \end{tabular}
\end{table*}%

\subsubsection{Counterfactual Questions}
Our approach uniquely integrates a causality framework, enabling predictions in hypothetical scenarios and addressing the counterfactual question outlined in Section~\ref{sec:introduction}.

In these experiments, we intervene on the $A\beta$ level in CSF or the age of the MCI group to predict counterfactual tabular variables and synthesize corresponding MRIs.

The MCI group is bifurcated into stable MCI (sMCI), who remain in the MCI stage, and progressive MCI (pMCI), who progress to AD. Identifying subgroup affiliation is crucial for timely treatment. In the ADNI test set, we examine 50 MCI patients, including 33 pMCI and 17 sMCI, across 87 and 36 data sessions, respectively.

\begin{figure}[!t]
\centering
\begin{minipage}[b]{.49\linewidth}
    \centering
    \subfloat[\scriptsize{Intervention}]{\label{fig:intervened_abeta}\includegraphics[width=0.475\linewidth]{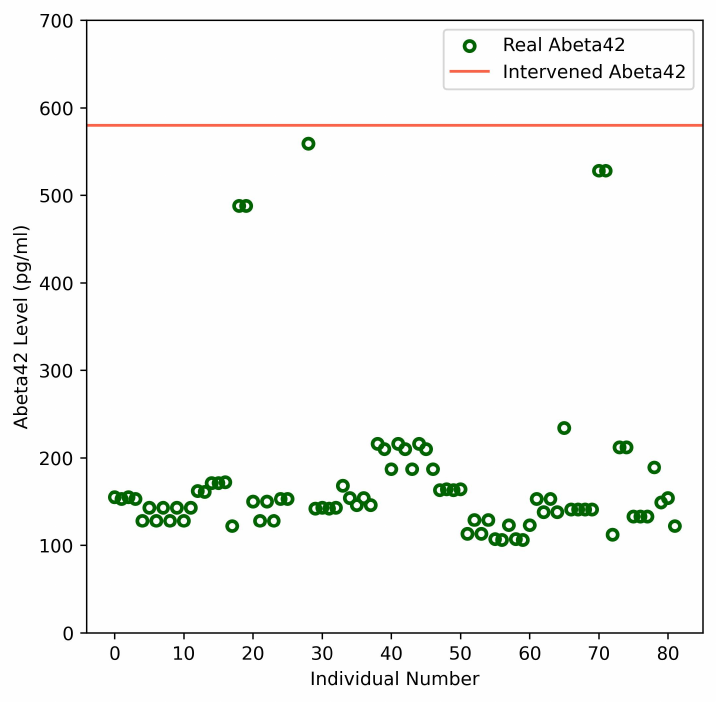}}
    \subfloat[\scriptsize{GMV}]{\label{fig:intervened_abeta_GMV}\includegraphics[width=0.485\linewidth]{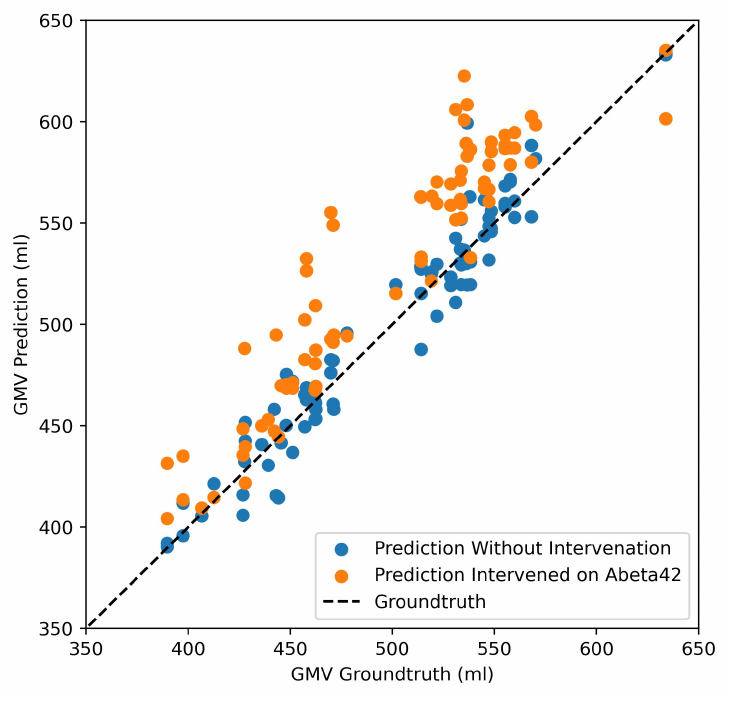}}
\end{minipage}
\begin{minipage}[b]{.49\linewidth}
    \centering
    \subfloat[\scriptsize{VV}]{\label{fig:intervened_abeta_VV}\includegraphics[width=0.485\linewidth]{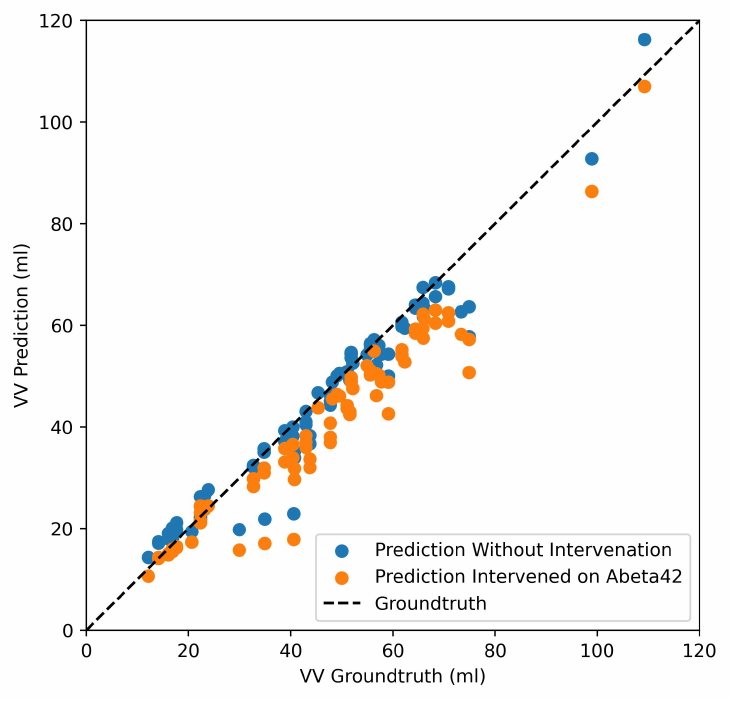}}
    \subfloat[\scriptsize{TIV}]{\label{fig:intervened_abeta_TIV}\includegraphics[width=0.495\linewidth]{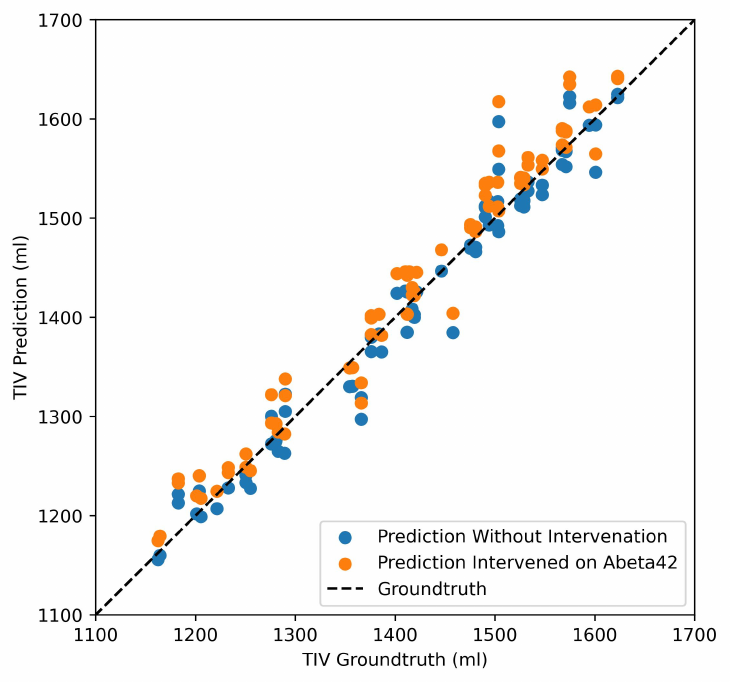}}
\end{minipage}
\caption{The illustration of (a) the $A\beta$ level in the subgroup pMCI and the intervened level (the mean level in sMCI) and the prediction of (b) GMV (c) VV, and (d) TIV with/without intervention of $A\beta$ and the ground truth. The y-axis is the prediction and the x-axis is the according ground truth, thus a data point below the dashed line indicates a prediction higher than the ground truth, vice versa.}
\end{figure}

Figures~\ref{fig:intervened_abeta_GMV}, \ref{fig:intervened_abeta_VV}, and \ref{fig:intervened_abeta_TIV} illustrate predictions of GMV, VV, and TIV when intervening on $A\beta$ levels in CSF for pMCI. Each individual is intervened to reach an $A\beta$ level of 474.41 pg/ml, the sMCI mean in ADNI. Figure~\ref{fig:intervened_abeta} reveals that most pMCI individuals have lower $A\beta$ levels than this mean (the red line). This suggests that increasing $A\beta$ in CSF may lead to larger GMV and TIV and smaller VV.

For counterfactual predictions, we use the same classifier as in Section~\ref{sec:downstream}. Of the 33 pMCI patients progressing to AD, 24 are correctly identified as pMCI using factual synthesized MRI. However, this number drops to 15 with counterfactual MRI intervened at the $A\beta$ level, suggesting that such interventions might slow AD progression. However, this insight is preliminary and requires clinical validation.

\begin{figure}[!t]
\centering\includegraphics[width=0.95\linewidth]{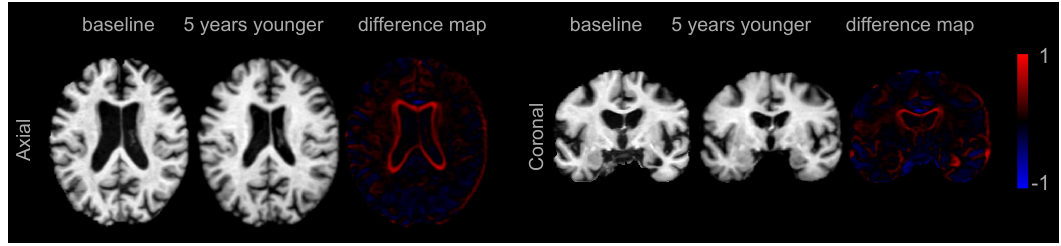}
\caption{The images display the MRI of a pMCI patient at the MCI stage alongside a synthesized MRI created by intervening on the age to appear five years younger. The first row presents the coronal slice, and the second row depicts the axial slice.}\label{fig:5yearsyounger}
\end{figure}

Figure~\ref{fig:5yearsyounger} illustrates how a pMCI patient's brain image might look if they were five years younger. We compute the difference map as ${I}_{diff} = {I}_{intervened} - {I}_{original}$, and the counterfactual MRI reveals a smaller ventricle and greater grey matter retention, aligning with current research~\cite{age_and_GM, age_and_VV1, age_and_VV2, age_and_VV3}. This approach can provide insights into past brain atrophy rates and assist clinicians in future progression predictions.

\section{Conclusion} \label{sec:conc}

In this paper, we introduce the Tabular-Visual Causal Graph (TVCG) model for the Causal Longitudinal Image Synthesis (CLIS) task. By integrating causality and longitudinal analysis into image synthesis, TVCG avoids spurious correlations and surpasses previous methods in performance. Besides, the synthesized images of TVCG also show significant promise in clinical AD characterization. 

A limitation of our model is the lack of modalities like PET and DTI, which could provide additional, crucial information that are not available in T1 MRIs. Future work should explore incorporating these modalities and examining their causal relationships. Additionally, our data-driven causal model building necessitates large datasets, so finding a more data-efficient construction approach is a key area for future research. All these further endeavors will guide us in evaluating CLIS’s practicality in real-world clinical contexts, encompassing AD and various other potential applications.




\bibliographystyle{ieeetr}
\bibliography{ref}

\begin{IEEEbiography}[{\includegraphics[width=1in,height=1.25in,clip,keepaspectratio]{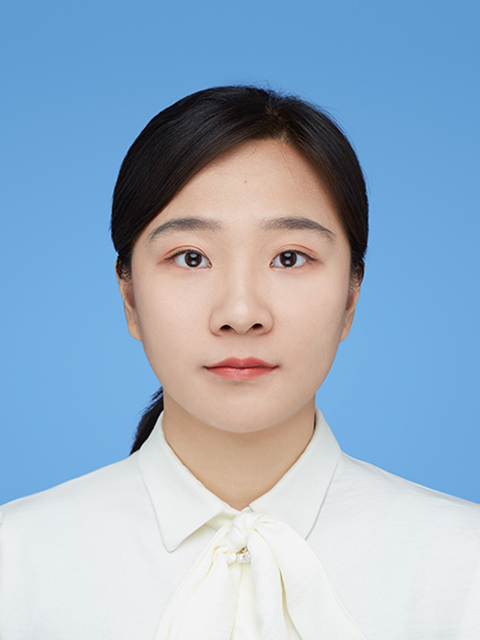}}]{Yujia Li}
received the B.E. degree from Tsinghua University in 2021. She is currently pursuing the Ph.D. degree from the and the M.S. degree in computer science from the Institute of Computing Technology (ICT), University of Chinese Academy of Sciences (UCAS). Her research interests include computer vision, machine learning and image processing, with applications to biomedical engineering.
\end{IEEEbiography}

\begin{IEEEbiography}[{\includegraphics[width=1in,height=1.25in,clip,keepaspectratio]{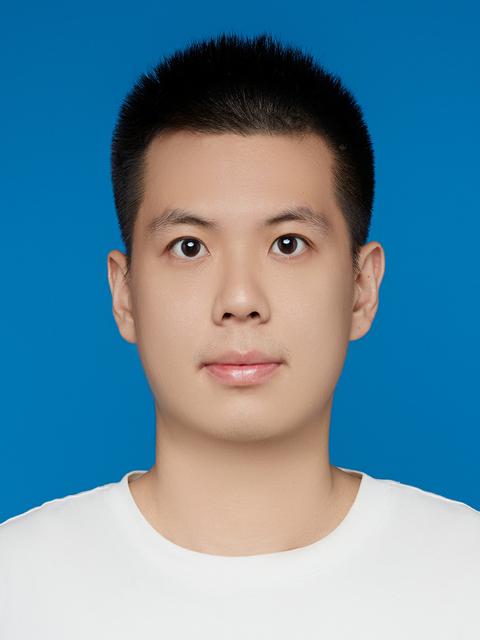}}]{Han Li}
received the B.E. degree from Henan Polytechnic University (HPU) in 2016, and the M.S. degree in computer science from the Institute of Computing Technology (ICT), University of Chinese Academy of Sciences (UCAS) in 2021. He is currently pursuing the Ph.D. degree from the School of Biomedical Engineering \& Suzhou Institute for Advanced Research, University of Science and Technology of China (USTC). His research interests include computer vision, machine learning and image processing, with applications to biomedical engineering.
\end{IEEEbiography}

\begin{IEEEbiography}
[{\includegraphics[width=1in,height=1.25in,clip,keepaspectratio]{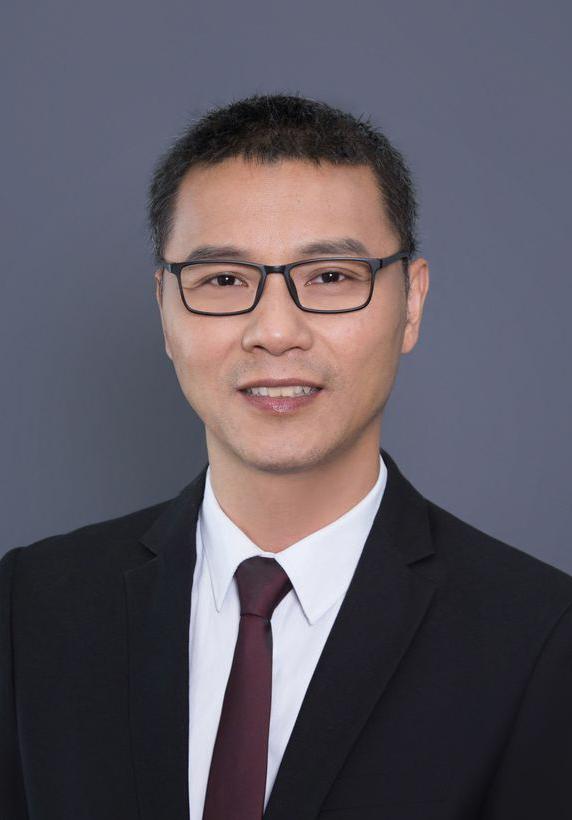}}]{Prof. S. Kevin Zhou} (Fellow, IEEE) obtained his Ph.D degree from the University of Maryland, College Park. Currently, he is a distinguished professor and founding executive dean of the School of Biomedical Engineering, Suzhou Institute for Advanced Research, University of Science and Technology of China (USTC), and an adjunct professor at the Institute of Computing Technology, Chinese Academy of Sciences. He directs the Center for Medical Imaging, Robotics, Analytic Computing and Learning (MIRACLE). Prior to this, he was a principal expert and a senior R\&D director at Siemens Healthcare Research. Dr. Zhou has published 260+ book chapters and peer-reviewed journal and conference papers, registered 140+ granted patents, written three research monographs, and edited three books. The most recent book he led the edition is entitled ``Handbook of Medical Image Computing and Computer Assisted Intervention, SK Zhou, D Rueckert, G Fichtinger (Eds.)" and the book he coauthored most recently is entitled ``Deep Network Design for Medical Image Computing, H Liao, SK Zhou, J Luo". He has won multiple awards including R\&D 100 Award (Oscar of Invention), Siemens Inventor of the Year, UMD ECE Distinguished Alumni Award, BMEF Editor of the Year, and Finalist Paper for MICCAI Young Scientist Award (twice). He has been a program co-chair for MICCAI2020, and an associate editor for IEEE Trans. Medical Imaging, IEEE Trans. Pattern Analysis Machine Intelligence, Medical Image Analysis, and an area chair for AAAI, CVPR, ICCV, MICCAI, and NeurIPS. He has been elected as a treasurer and board member of the MICCAI Society, an advisory board member of MONAI (Medical Open Network for AI), and a fellow of AIMBE, IAMBE, IEEE, MICCAI, and NAI.
\end{IEEEbiography}

\end{document}